\documentclass{article}
\usepackage{amsmath, amscd, amsfonts, amssymb, amsthm, tikz, times, setspace, srcltx}
\usepackage{color}

    \newtheorem{theorem}{Theorem}[section]
    \newtheorem{corollary}[theorem]{Corollary}
    \newtheorem{lemma}[theorem]{Lemma}
    {\bf}{\rm}
    \numberwithin{remarks}{section}
    \newtheorem{proposition}[theorem]{Proposition}
    \newtheorem{definition}{Definition}
    \numberwithin{theorem}{section} \numberwithin{definition}{section}

\def\be{\begin{equation}}
\def\ee{\end{equation}}

\def\Z{\mathbb{Z}}
\def\Y{\mathbb{Y}}
\def\N{\mathbb{N}}
\def\R{\mathbb{R}}
\def\D{\mathbb{D}}
\def\K{{K}}

\def\F{\mathcal{F}}
\def\L{\mathcal{L}}

\def\GG{\mathcal{G}}
\def\r{\mathcal{R}}

\def\H{{\mathcal S}_{\lambda}}

\newcommand{\dd}{d} 
\DeclareMathOperator{\const}{const}
\DeclareMathOperator{\dist}{dist}

\def\1{1}
\def\eps{\varepsilon}
\def\P#1{ \mathbb{P} \left( {#1} \right) }

\begin{document}
\title{Symmetry breaking in quasi-1D Coulomb systems}
\author{Michael Aizenman$^{(a)}$
\footnote{Supported in part by NSF grant DMS-0602360}
   \quad  Sabine Jansen$^{(b)}$
   \footnote{Supported in part by DFG Forschergruppe 718 ``Analysis and Stochastics in Complex Physical Systems",  \newline
\hspace{5cm} NSF grant PHY-0652854 and a Feodor Lynen research fellowship of the Alexander von Humboldt-Stiftung}
   \quad   Paul Jung$^{(c)}$
   \footnote{Supported in part by Sogang University research grant 200910039}
\\[2ex]
{\normalsize  $^{(a)}$ Departments of Physics and Mathematics,
  Princeton University,  Princeton NJ 08544, USA  } \\[1ex]
{\normalsize $^ {(b)}$  Weierstrass Institute for Applied Analysis
  and Stochastics, Mohrenstr. 39, 10117 Berlin, Germany}\\[1ex]
{\normalsize $^ {(c)}$  Department of Mathematics,
  Sogang University, 121-742 Seoul, South Korea}\\[1ex]
   }
\date{August 20, 2010}

\maketitle
\abstract{Quasi one-dimensional systems are systems of particles in domains
which are of infinite extent in one direction and of uniformly bounded size in all other directions, e.g. on a cylinder of infinite length.
 The main result proven here is that for such particle systems with Coulomb interactions and neutralizing background, the so-called ``jellium'', at any temperature and at any finite-strip width there is translation symmetry breaking.   This extends the previous result on Laughlin states in thin, two-dimensional strips by Jansen, Lieb and Seiler (2009).
The structural argument  which is used here bypasses the question of whether the translation  symmetry breaking is manifest already at the level of the one particle density function.  It  is akin to that employed by  Aizenman and Martin (1980) for a similar statement concerning symmetry breaking at all temperatures in strictly one-dimensional Coulomb systems.  The extension is enabled through bounds which establish tightness of finite-volume charge fluctuations.
\\

}
\newpage
\tableofcontents

\section{Introduction}
\nocite{*}

 In this work we investigate symmetry breaking in classical quasi one-dimensional ``jellium'',  that is particle systems with Coulomb repulsion and attractive neutralizing background (also known as ``one-component plasma''), and in quantum systems whose states may be described by such ensembles.    The  particles are of equal  charge $-q$, and move in domains which are of infinite extent in one direction and of uniformly bounded size in all
 other directions, e.g. cylinder or tube of infinite length and a finite, uniform,  cross-section.   The Coulomb potential is the solution to a
 Poisson equation with Neumann or periodic boundary conditions in the confined directions.
It corresponds to the situation where not only the particles but also  the electric field  is confined to the tube.

Our main result is that such systems display translational symmetry breaking in the long direction, e.g., the cylinder axis, which is denoted here
by $x$. This generalizes previously known results in one and two dimensions. The proof is by a structural argument which is not limited to low temperatures or small tube cross sections.

For the one-dimensional jellium  this phenomenon was shown in \cite{K,BL,AM}. The periodicity is related to the possible formation of a ``Wigner lattice''.   Roughly, the Coulomb interaction leads to the suppression of large scale deviations from neutrality.   Each  particle, as ranked by
the $x$ coordinate,  fluctuates with only a bounded deviation from the lattice value of $x$ corresponding to its rank.

In two dimensions, with periodic boundary conditions in the confined direction, symmetry breaking is known for some special cases corresponding
to even-integer values of the so-called ``plasma parameter'' $\beta q^2$ (with $\beta$ the inverse temperature).  When $\beta q^2=2$, the model is explicitly solvable and periodicity was shown in \cite{CFS}. A proof of symmetry breaking for other even-integer values $\beta q^2 = 2p$
and sufficiently small values of the
strip width (compared to $(\text{density})^{-1/2}$)  was given in~\cite{JLS}, where the focus was on the Laughlin states in cylindrical geometry.   Numerical results may be found in~\cite{swk}.

The case of even-integer values of the plasma parameter is of an additional interest, and further tools are available for it, as it relates to   \emph{Laughlin's wave function}~\cite{lau} which  is frequently used as an approximate ground state for electrons in the context of the fractional quantum Hall effect.
The function models the state of an electron gas whose filling factor
(a quantity related to the electron density) is a simple fraction $1/p$ with $p$ an integer.
The function's modulus squared is proportional to the Boltzmann
weight of a classical one-component plasma: $ |\Psi|^2 \propto \exp(-\beta U)$.
The filling factor of an electron gas and the plasma parameter of a classical Coulomb system are related by $\beta q^2 = 2p$.
The solvable case $\beta q^2 = 2$  corresponds to $p=1$ (filled lowest Landau level), which models non-interacting fermions.
The integrality of $p$ enters the proof in \cite{JLS} in a crucial way, allowing to expand $p$-th powers of polynomial into monomials.

The proof given here follows an altogether different approach, along the lines of \cite{AM, AGL}.
A key point is that due to the strong tendency of Coulomb systems to maintain bulk neutrality an interesting phenomenon occurs in one-dimensional and quasi one-dimensional Coulomb systems : the total charge in cylinders of arbitrary length is of bounded variance.  Even in the infinite-volume limit the question ``what is the total charge at points with   $x\le x_0$?''  has a well defined answer (denoted here by ${q} K(x,\omega)$).    As was pointed out in \cite{AGL} that fact in itself implies symmetry breaking, through the long-range correlations in the values of the phase $e^{i 2\pi K(x,\omega)}$, or equivalently through the fractional part of the total charge below $x_0$.
This approach to symmetry breaking was developed in the earlier work on strictly one-dimensional Coulomb systems~\cite{AM}, in which case $qK(x,\omega)$ 
yields the electric field at $x$.  In \cite{AGL} this was extended into  a general criterion that in one dimension, ``tightness'' of charge fluctuations implies symmetry breaking.   To make this argument  applicable in our setting one needs to first establish the tightness estimates.  These are somewhat more involved than in the strictly one-dimensional case.   Curiously, in both~\cite{JLS} and here symmetry breaking is explained  through a combination of analytical and topological arguments.  However, these seem to be of somewhat different nature in these two works.

The paper's outline  is as follows.   The model is introduced in  Section \ref{jellium}.   The main results are stated in  Section~\ref{sec:main results}; that includes the statement of  Theorem \ref{thm:theta} and its Corollary~\ref{coro:TSB} which
addresses the symmetry breaking in the model's infinite-volume Gibbs states.  The results are enabled by  a pair of estimates which play an essential role: Theorems~\ref{thm:tightness} and \ref{thm:average}.    We then turn to the mathematical framing of the particle-excess function $K(x,\omega)$, which it is convenient to view alternatively as a random element of a Skorokhod space, and as a cocycle in the sense of dynamical systems.  These terms are discussed in Section~\ref{sec:charge_excess} and used in Section~\ref{sec:condproof} for a conditional proof of symmetry breaking, assuming the tightness bounds.  The proof of the latter involves a different set of considerations.  These are outlined in Section~\ref{1Dsection}, which reviews the corresponding question in one dimension.  Finally, the proof of the enabling bounds is spelled out in Sections~\ref{sec:tightness}, where we establish tightness of the distribution of $K(x,\omega)$ at fixed $x$, and
 in Section~\ref{sec:averages} where it is shown that the volume-averages of $K(x,\omega)$ tend to null.  We end with a few additional remarks in Section~\ref{sec:discussion}.

\section{Coulomb interaction in quasi one-dimensional systems} \label{jellium}

\subsection{The potential function}

The quasi one-dimensional systems considered here are systems of
particles which, along with the electric field they generate, are
confined to a tubular region of the form $\mathcal{T}=\R\times \D$
where $\D$ is a  compact subset of $\R^k$, possibly with some
periodic boundary conditions.  The precise technical assumption is
spelled out below.
In the simplest example $\mathcal{T}$ is a strip whose
cross-section  is $\D=[0,W]$ with periodic boundary conditions.

Points on $\mathcal{T}$ will be denoted $z=(x,y)$ with $x\in \R$ and $y\in \D$.
For simplicity we denote the volume-form on  $\D$ by $dy$  and its total
measure by $W=\int_\D dy$.

The Coulomb potential between two points is given by a symmetric function,
$V(z,z') = V(z',z) $, which satisfies:
\be \label{delta}
-\Delta V(z,z') \ = \  \delta(z-z')
\ee
for $-\Delta = -\frac{\partial \, ^2}{\partial x\, ^2}  - \Delta_\D$, with
$\Delta_\D$ the Laplacian  on $\D$ which is taken here to be defined  with either periodic or Neumann boundary conditions.

Since  $\mathcal{T}$ has locally the structure of $\R^d$, with $d=1+k$, the
short distance behavior of the potential at interior points of $\D$ is
\begin{equation}
V(z,z') \  \approx \  \begin{cases}[(d-2)C_d]^{-1}\dist (z,z')^{-(d-2)} &
\mbox{for  } d\neq 2\\[1ex]
-(2\pi)^{-1} \ln\dist (z,z')  & \mbox{for  }  d=2 \end{cases}    \quad  .
\end{equation}
Yet, at long distances $V(z,z')$ behaves as a one-dimensional Coulomb
potential:
\begin{equation}
V(z,z') \  \approx \   -|x-x'|/(2W)   \, ,
\end{equation}
in a sense which we shall now make more explicit.

In the example of the 2D periodic strip  it is convenient to use  the complex
notation: $z=x+iy$,  in terms of which
\be
	V(z,z')\  =\  -(2\pi)^{-1}\log|2\sinh(\pi (z-z')/W)| \, .
\ee
This function is clearly periodic in $y={\rm{Im}}  \ z$ and  harmonic throughout
the (periodic) strip except at $z=0$,  and it can be easily seen to satisfy
\eqref{delta}.  At short distances $V(z,z')$  behaves as the two-dimensional
Coulumb potential, with  logarithmic divergence, but its long distance behavior
is close to that in one dimension, and better described by the decomposition:
\be \label{potential_split}
	V(z,z')    \ =\     -|x-x'| /(2W)  \ +  \  V_2(y,y'; |x-x'|)
\ee
with  the correction to the linear term given by
\begin{eqnarray}\label{totalenergy}
 V_2(y,y'; |x-x'|) \  =\  -(2\pi)^{-1}\log|1-e^{-2\pi (|x-x'|+i|y-y'|)/W}|  \, ,
\end{eqnarray}
which decays  exponentially in  $|x-x'|$.

In the more general case the potential admits the eigenfunction expansion:
\begin{eqnarray} \label{kernel_expansion}
V(z,z')  &  = &
  -|x-x'|/(2W)  \ + \    \sum_{n\ge 1}   (2\sqrt{E_n})^{-1}\, e^{-|x-x'| \sqrt{E_n}
}    \,  \overline{\varphi_n(y)}\varphi_n(y')  \nonumber  \\[2ex]
      &  =: &   -|x-x'|/(2W)  \ + \  V_2(y,y';|x-x'|) \, .
\end{eqnarray}
in terms of the eigenfunctions of $\Delta$:  $-\Delta_\D \varphi_n(y) = E_n
\varphi_n(y)$.

We may now state our assumptions on $\D$, which are:
\begin{enumerate}
\item  The  correspondingly  periodic / Neumann   Laplacian $\Delta_\D$ is a self
adjoint operator  with  a non-degenerate ground state ($\varphi(y) = W^{-1/2}$)
and compact resolvent.
\item  The Coulomb potential in $\mathcal{T}$ is of the form
\eqref{kernel_expansion} whose second term averages to zero over $\D$:
\be \label{average}
	\int_\D V_2(y,y';|x-x'|) \, dy \ = \ 0 \, ,
\ee
for any $x, x' \in \R $ and $ y'\in \D$.
\item   The term $V_2$ admits bounds of the form
\be \label{lowerbound}
	 V_2(y,y';|x|) \ \ge \  - g(|x|)
\ee
and for each $\delta>0$ there exists $c(\delta)>0$ such that
\be \label{upperbound}
	 V_2(y,y';|x|) \ \le \  c g(|x|) \ \text{ for all }\ |x|\ge\eps
\ee 
for $g(x)$ a positive decreasing function on $[0,\infty)$,  which   satisfies the
`finite energy condition':
\be \int_0^\infty  x\,  g(x)\,  dx \   < \   \infty \,  .
\ee
\end{enumerate}

\begin{figure}[h]\label{fig:flux}
	\begin{center}
		\includegraphics[scale=.4] {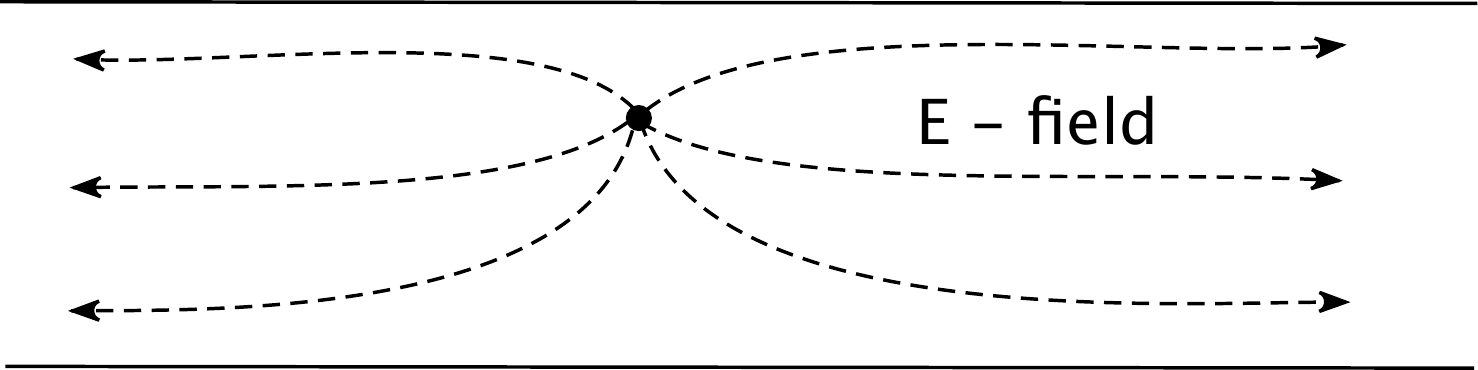}
		\begin{caption} {\footnotesize
			Flux lines of the potential with Neumann  boundary conditions.}
		\end{caption}
	\end{center}
\end{figure}

The boundary conditions ensure that the electric flux lines (i.e., lines tangent to
$\nabla V$) do not leave the tube, see Fig.~\ref{fig:flux}.

The averaging condition  \eqref{average} is equivalent to saying that the
interaction of a uniform slab with a point charge does not depend on the charge's
position within the tube's cross section.  In terms of the eigenfunction
expansion this is implied by the constancy of the ground state
$\varphi_0$, to which all other eigenstates $\varphi_n$ are orthogonal.    The
decay in $|x|$ of the individual terms is due to the spectral gap.   However, one
has to control the sum in \eqref{kernel_expansion}, whose terms need not be
uniformly bounded (in $L^\infty$).   We owe to Rupert Frank  the
comment that conditions \eqref{lowerbound} and \eqref{upperbound} hold for compact Lipschitz
domains.
In $\R^2$, this class includes compact domains with piecewise differentiable
boundary, which may exhibit discontinuities in the tangent's direction but no
``horn singularities" of vanishing angle.
For bounded domains $\D$ with smooth boundary, these conditions hold with
$g(x)$ of exponential decay.

\subsection{Quasi one-dimensional jellium}\label{sec:q1D jellium}

Jellium in  a finite segment, corresponding to
$$ \mathcal{T}_{[L_1,L_2]}=[L_1,L_2]\times \D \quad \text{with}\ \ L_1,L_2\in\R \, , $$ 
consists of a collection of $N \approx \rho(L_2-L_1) W$ particles of charge $(-q)$ each,
with
$\rho$  the mean number of particles per volume,
moving in a neutralizing background of homogeneous charge density $q \rho$.

We shall denote a particle configuration by $\omega=(z_1,\ldots,z_{N})$, with
$z_j = (x_j,y_j) \in \mathcal{T}_{[L_1,L_2]}$ and
the labeling chosen in the increasing order of the $x$-coordinates.
Thus, the  configuration space is  $\Omega_{[L_1,L_2]}^{(N)}\ =\  \triangle(N,[L_1,L_2])
\times \D^N$ with
$\triangle(N,[L_1,L_2])$ the simplex
$\{x\in \R^N \, :\, L_1 < x_1<\cdots<x_{N}< L_2\}$.

Using \eqref{kernel_expansion} and \eqref{average}, one gets for the system's energy, discarding the finite self-interaction of the fixed background:
\begin{eqnarray}\label{energy}
        \nonumber U(z_1,\ldots,z_N)
        &=&\sum_{1\le j<k\le N} q^2 V(z_j,z_k)-\sum_{j=1}^{N}
                         q^2\rho \int_{\mathcal{T}_{[L_1,L_2]}} V(z_j,z) \,  \dd z\\[2ex]
        &=& \sum_{1\le j<k\le N} q^2 V(z_j,z_k)       + \frac{1}{2}q^2 \rho
                \sum_{j=1}^{N}\Bigl(x_j - \frac{L_1+L_2}{2} \Bigr)^2+\text{Const}(N,L)\,.
\end{eqnarray}
with $\dd z = \dd x\thinspace \dd y$ the Lebesgue measure on $\mathcal{T}_{[L_1,L_2]}$  and $L=L_2-L_1$.

The jellium's Gibbs equilibrium state, at the inverse temperature $\beta \equiv
\frac{1}{kT}$, is the probability measure on $\Omega^{(N)}_{[L_1,L_2]} $:
\be
 \  e^{-\beta U(\omega)} \, d\omega_{[L_1,L_2]}
/Z(\beta, N, L)
\ee
where $d\omega_{[L_1,L_2]}$ denotes the product Lebesgue measure on
$\Omega^{(N)}_{[L_1,L_2]} $.
The normalizing factor $Z(\beta, N,L)$ is  the finite volume partition function.

One may note that for  $\D=[0,W]$ with the
periodic boundary conditions and $\beta q^2= 2p$, with $p$ an integer, this
measure coincides with the Laughlin states, $|\Psi_p(z_1,...,z_N)|^2$, which were
considered in \cite{thou}.

\paragraph{Dimensionless parameters and choice of units}

One could, for convenience, rescale the length units (uniformly in all directions) so that the particle density is  $\rho=1$, and, since $q$ affects the state only in the combination $\beta q^2$, absorb $q^2$ in the units of  $\beta$  thereby setting $q=1$.   With this choice one is still left with dependency on two parameters, $\beta$ and $W$.  We shall follow this choice when referring to constants which appear in bounds  throughout the paper as $C(\beta,W)$.  However, in other places we shall leave the $\rho$ and $q$ dependence explicit.

An explicitly  relevant length scale for us is
\be \label{ell}
\lambda \ :=\  (\rho W)^{-1}  \, .
\ee
This is the length of a cylindrical cell in which the mean number of particles is one.   It is also the basic scale for the translation symmetry breaking proven here.

 It may be added that for dimensionless parameters of the model one may chose $W^{1/(d-1)}/\lambda = \rho W^{d/(d-1)}$ and the ``plasma parameter'', sometimes called the coupling constant, $\Gamma_d = \beta q^2 \rho^{(d-2)/d}$.

\section{Statement of the main results}\label{sec:main results}

\subsection{Symmetry breaking}

Our main result is the following statement concerning the infinite-volume limits of the finite volume Gibbs equilibrium states of the
quasi one-dimensional systems in the regions $\mathcal{T}_{[L_1,L_2]}$.

As is explained by the result, it is natural to take the boundaries of the finite regions as:
\be \label{eq:Lj}
L_1 \ =\ (- n_1-\theta ) \, \lambda \quad , \qquad  L_2\ =\  (n_2-\theta ) \,  \lambda
\ee
with $n_1, n_2 \in \N$, and $\theta \in [0,1)$.  The corresponding Gibbs equilibrium  measure for the neutral systems of $N = n_1 + n_2$ particles in such
domains will be denoted here
$\mu^{(\beta, W, \theta)}_{n_1,n_2}$.   For the purpose of convergence statements, these can be viewed as point processes on the common
space $\mathcal{T}=\R\times \D$.

\begin{theorem}  \label{thm:theta}
At all $\beta >0$  and $W>0$:
\begin{enumerate}
\item  For any $\theta \in [0,1)$ and any sequence of integer pairs $\{(n_1(j), n_2(j))\}$, with $n_1(j), n_2(j)  \to \infty$, there is
a subsequence for which $\mu^{(\beta, W, \theta)}_{n_1,n_2}$ converges to a limit,   in the sense of convergence of probability measures on
the space of configurations in $\mathcal{T}$ (i.e. of point processes).
 \item   Any two limiting measures which correspond to different values of $\theta$ are mutually singular.
 \item More explicitly:
 there exists a measurable function on the space of configurations,  such that for each $\theta \in [0,1)$:
 \be \Phi(\omega) \ = \ e^{i 2\pi \theta}
 \ee
almost surely with respect to each measure which is a limit of a sequence of finite-volume `$\theta$ Gibbs states'.

Furthermore, under the shifts (induced by $T_u : \, x  \mapsto  x+u$) and reflections (induced by ${\mathcal R}: \, x  \mapsto  -x)$:
\be  \label{eq:phimaps}
 \Phi(T_u\omega) =  \Phi(\omega)  \, e^{i 2\pi u /\lambda }\,  \qquad
 \mbox{and   \quad $\Phi({\mathcal R}\omega) = \bar \Phi(\omega)$\,.}
\ee
\end{enumerate}
\end{theorem}
This has the following elementary consequence:
\begin{corollary}[Symmetry breaking] \label{coro:TSB}
\mbox{}

\begin{enumerate}
\item  None of the limiting measures discussed above is invariant under the shifts $T_u$, with $u\notin \lambda  \Z$ (at $\lambda\equiv (W\rho)^{-1} )$.
 \item Except  for  the two cases $\theta = 0, 1/2$ (for which $\theta \equiv 1-\theta _{\mbox{mod. $1$}}$), the limiting measures are also not invariant under the reflection ${\mathcal R}$.
\end{enumerate}
\end{corollary}

The derivation of the above statements combines three elements:  ``statistical mechanical'' bounds on the distribution of the `particle-excess function', which are presented below,  dynamical systems' concepts of tight cocycles, and simple topological considerations.

A fundamental role in the analysis  is played by the `particle-excess' function $K(u)$, which expresses the difference (below $u$) of the total background charge times $q^{-1}$ and the number of particles:
\be
K(u,\omega) \ :=\
\begin{cases}
\frac{u-L_1}{\lambda}   - |\{ j \le N\, : \, x_j \le  u\} |  & \mbox{if $u\in [L_1,L_2]$} \\[2ex]
\qquad 0 & \mbox{if $u\in \R \backslash [L_1,L_2]$}
\end{cases}
\ee
with $|\cdot|$ the cardinality of the set (see Figure 2).  For a point process on the line it is rather exceptional that such a quantity has a good infinite-volume limit ($L_1\to -\infty$, $L_2\to \infty$).  When it does, translation symmetry breaking follows by the general argument of \cite{AGL}.

\begin{tikzpicture}[scale=1]
\draw (0,-5) node [right,text width=115mm]
   {\footnotesize Figure 2: A particle configuration on $[L_1,L_2]\times\D$ (here $\D=[-\pi,\pi], \rho=1$) along with its associated particle-excess function $K(x)$.};    \label{fig:K}

Draw thin grid lines with color 40
 \draw [step=0.5,thin,gray!55] (0,-1.571) grid (6.0,1.571);
 \draw [] (0,-1.571) node [below] {$L_1$} -- (6.0,-1.571) node [below] {$L_2$} -- (6.0,1.571) -- (0,1.571) ;
 \draw [] (0,-1.571) node [left] {$-\pi$} -- (0,1.571) node [left] {$\pi$};
 \draw [semithick,red] (.8,.9) circle (.05) ++(.15,-1.65) circle (.05) ++(.12,.89) circle (.05)
 ++(.73,1.2) circle (.05) ++(.25,-1.75) circle (.05) ++(.71,1.1) circle (.05)
 ++(.15,-1.85) circle (.05) ++(.89,1.4) circle (.05) ++(.3,.97) circle (.05)
 ++(.6,-1.9) circle (.05) ++(.5,.92) circle (.05) ++(.6,-1.63) circle (.05);
 \path (0,-3.5) coordinate (origin);
 \draw [->] (origin) node [below, xshift=-.2cm] {$L_1$} --++(6.0,0) node [below, xshift=.2cm] {$L_2$};
 \draw [->] (0,-4.5) -- ++(0,2) node [left] {$K(x)$};
 \draw [step=0.5,thin,gray!40] (0,-4.49) grid (6.0,-2.51);
 \draw [blue] (origin)--++(.8,.8)--++(0,-.5)--++(.15,.15)--++(0,-.5)--++(.12,.12)--++(0,-.5)
 --++(.73,.73)--++(0,-.5)--++(.25,.25)--++(0,-.5)--++(.71,.71)--++(0,-.5)
 --++(.15,.15)--++(0,-.5)--++(.89,.89)--++(0,-.5)--++(.3,.3)--++(0,-.5)
 --++(.6,.6)--++(0,-.5)--++(.5,.5)--++(0,-.5)--++(.6,.6)--++(0,-.5)--(6,-3.5);
 \draw [red] (.8,-3.5) circle (.03) ++(.15,0) circle (.03) ++(.12,0) circle (.03)
 ++(.73,0) circle (.03) ++(.25,0) circle (.03) ++(.71,0) circle (.03)
 ++(.15,0) circle (.03) ++(.89,0) circle (.03) ++(.3,0) circle (.03)
 ++(.6,0) circle (.03) ++(.5,0) circle (.03) ++(.6,0) circle (.03);
 \end{tikzpicture}

\subsection{Bounds on the distribution of the particle-excess function}

The results stated in Theorem~\ref{thm:theta}  are enabled by the following auxiliary bounds.  The first bound expresses the tightness of the distribution of the particle-excess function.
\begin{theorem}[Tightness bound] \label{thm:tightness}
For each $\beta >0$  and $W>0$, the following bound holds for all
 $n_1 , n_2 \in \Z$, $\theta\in [0,1)$, and $x\in [L_1, L_2]$  (with $L_j$ defined in \eqref{eq:Lj} ),
\be  \label{tighness}
		\mu^{(\beta, W, \theta)}_{[n_1,n_2]} \left (  \left \{   |K(x;\omega)| \ge \gamma   \right \}
		\right )   \  \le  \   C(\beta,W)\,  e^{-A(\beta,W)
		\gamma^2}
\ee
 at some
$C(\beta,W) < \infty $ and $A(\beta,W)>0$.
\end{theorem}

This bound can be made intuitive by noting that the Coulomb systems' energy  can be presented as $\frac{q^2\rho}{2} \int K(x)^2 dx$ plus the
short-range interaction corresponding to $V_2$.
In fact, a stronger statement is valid, with the exponent in \eqref{tighness}  replaced by
$-A(\beta,W)|\gamma|^3$.  The bound asserted in \eqref{tighness}  applies also to the two component Coulomb system.   It is stated here in
this weaker form in order to make the discussion of its implications applicable to also such systems.

The second bound expresses the asymptotic vanishing of the translation averages of $K(x,\omega)$:

\begin{theorem} [The vanishing of $K$'s volume-averages] \label{thm:average}
For each $\beta,W, \delta >0$, there exist $\widetilde C=\widetilde C(\beta, W, \delta) < \infty $ and $\alpha = \alpha (\beta, W)>0$ with which
\be  \label{eq:average}
\mu^{(\beta, W, \theta)}_{[n_1,n_2]} \left (  \left \{  \frac{1}{{r}} \left |\int_0^{r} K(x+u;\omega) \, du \right | \ge \delta   \right \}  \right )
\  \le  \   \widetilde C \,  e^{-\alpha  \  \delta^2  \  {r} }
\ee
for any $n_1 , n_2 \in \Z$,  $x\in \R$, and any ${r}>0$.
\end{theorem}
Since $K(x,\omega)=0$ for $x\notin  [L_1, L_2]$, it suffices to verify   \eqref{eq:average}  for all $x\in   [L_1, L_2]$ and ${r}\in [0, L_2- x]$.
To simplify the notation, we shall sometimes write $\mathbb{P}$ instead of $\mu_{[n_1,n_2]}^{(\beta,W,\theta)}$.

Before presenting the derivation of these estimates (Sections~\ref{replacementsection} and \ref{sec:averages}), we shall show how they imply symmetry breaking, i.e. give a conditional proof of Theorem~\ref{thm:theta}.   First however, let us frame the discussion of the function $K(x;\omega)$  within a convenient setup.

\section{Basic properties of the particle-excess function}  \label{sec:charge_excess}

\subsection{A convenient topological setup}

It is convenient to view the particle-excess function $K(x;\omega)$ as a random variable with values ranging over a  subset of the   Skorokhod space,  which we denote by $\H$,  of  functions on $\R$  which
\begin{enumerate}
\item  have the c\`adl\`ag  property (continuity  from the right, and existence of limits from the left),
\item  have only integer valued discontinuities,
\item  are piecewise differentiable with a constant derivative,  given by $\rho W\equiv 1/\lambda$.
\end{enumerate}
It may be noted that for elements of $\H$ the c\`adl\`ag continuity modulus ${\textit{w}}_K(\delta)$, which expresses the maximal variation of  $K(x;\omega)$ between pairs of points at distance $\delta$ omitting a finite number of  discontinuities~\cite{cadlag},
takes the non-fluctuating  values:
\be  \label{modulus}   {\textit{w}}_K(\delta) \ = \ \delta/\lambda  \,  ,  \quad \mbox{ for all $\delta>0$ and $ K\in \H $.}\,
\ee

We shall make use of the following two observations, which are simple consequences of standard arguments.
\begin{lemma} \label{lem:tigness}
A sufficient condition for tightness of a family of probability measures supported on $\H$ is the
tightness with respect to this family of $\sup_{x\in \R} |K(x;\omega)/F(x)|$, for some continuous function $F(x)$ which diverges at $\infty$.
\end{lemma}
The proof is by a standard argument (which uses the Arzel\`a-Ascoli theorem), and will be omitted here.  Let us however note that the standard criterion for tightness of probability measures on the Skorokhod space requires as a second condition also the vanishing in probability of the continuity modulus $ {\textit{w}}_K(\delta) $  for $\delta \to 0$.  That, however, is directly implied for functions in $\H$  by~\eqref{modulus}.

\begin{lemma}
For $x\in \R$ and $K\in \H$ , the following `phase functional'
\be
\Phi(x,K) \ := \ e^{i 2\pi K(x)}
\ee
is continuous in $K$ with respect to the Skorokhod topology (restricted to $\H$).
\end{lemma}
The point here is that while the evaluation function $K\mapsto K(x)$ itself is not continuous, since the location of the jump discontinuities may change, the above phase is not affected by the location of the jumps when they are by integer amounts.

\subsection{The charge cocycle }

The total charge in an interval $(0,u]$, that is:
\be Q(u; \omega)   := \frac{q u}{\lambda}  \ - \ q |\{j: \, x_j \in (0,u] \} |  ,
\ee
can  be expressed as the difference:
\be \label{coboundary}
  Q(u; \omega)  \ = q\ K(0; T_u \omega) - q K(0; \omega)  \, .
\ee
In the  dynamical systems terminology, $ Q(u; \omega) $  is a cocycle under the action on $\H$ by the group of shifts.  That is, it transforms as:
\be \label{cocycle}
Q(u+s; \omega)\ = \  Q(u; T_s \omega) + Q(s; \omega)  \, .
\ee
Equation \eqref{coboundary}, states that this  cocycle is the coboundary of $K$, as long as the latter is well defined (a point which is not to be taken for granted in the infinite-volume limit).

For background it may be of relevance to recall the following general principle.  (To avoid excessive notation, we do not change here the notation from the specific to the general.)

\begin{proposition} [K. Schmidt (77)]  \label{prop:cocycle}

A quantity $Q(x; \omega)$ which transforms as \eqref{cocycle} under the action of measure preserving transformations $\{T_u\}_{u\in \lambda \Z}$ admits a representation similar to \eqref{coboundary},  in terms of a {measurable function $K_0(\omega)$},
\emph{if and only if}  the  collection of variables  $\{Q(u; \omega)\}_{u\in \lambda \Z}$ is  tight.
\end{proposition}

Tightness means in this context that the following bound holds uniformly
for $u\in \lambda \Z$
\be
\P{|Q(u; \omega)|  \ge t} \ \le \ p(t)
\ee
with some $ p(t)$ which vanishes for $t \to \infty$.
The less  elementary part of the proposition concerns the  \emph{`only if'} direction, for which a constructive argument can be provided~\cite{schmidt} (as discussed  also in  \cite{AGL}).

While the above proposition sheds light on our discussion, we shall bypass here the requirement of shift invariance by making use of the additional information given by Theorem~\ref{thm:average}.

\section{Conditional proof of Theorem~\ref{thm:theta}}   \label{sec:condproof}

Since the derivation of the enabling Theorems~\ref{thm:tightness} and \ref{thm:average}, which is presented (independently) in Sections~\ref{replacementsection} and \ref{sec:averages}, would take the discussion into a different arena than the one introduced above, let us first present their implication for our main results.

\begin{proof}[Proof of Theorem~\ref{thm:theta} -- assuming Theorems~\ref{thm:tightness} and \ref{thm:average}]$\\$

\noindent{\bf Existence of limits for subsequences}

Our strategy is to first discuss the question of convergence of the probability distributions at the level of $\H$, that is of the distribution of the random particle-excess function $K(x,\omega)$, which takes values in that Skorokhod space.
Since
\be
\sup_{x\in [n,n+1]} |K(x;\omega)| \ \le \   \max \{ |K(n;\omega)|,|K(n+1;\omega)| \} + \rho W
\ee
the bound of Theorem~\ref{thm:tightness} yields:
\be
\mu^{(\beta, W, \theta)}_{[L_1,L_2]} \left (  \left \{  \sup_{x\in\R} \frac{|K(x;\omega)|}{\sqrt{\ln (2+|x|)}} \ge t   \right \}
		\right )   \  \le  \   2 \, C(\beta,W)\, \sum_{n\in \N}  e^{-A(\beta,W)
		[t \sqrt{\ln n} -\rho W]^2}
\ee
where the upper bound is finite when $t^2 A(\beta,W) >1$ and vanishes, uniformly in
$\{\theta, L_1,L_2\} $,  for $t\to \infty$.
By Lemma~\ref{lem:tigness} this implies tightness of the probability measures  on $\H$ which are induced by  the Gibbs measures
$\mu^{(\beta, W, \theta)}$.  Hence, for every sequence of such measures there  exists a subsequence for which the induced probability measures on   $\H$ converge.
Since the charge configuration in any finite interval $I\subset \R$ is a continuous function of $K$ (in the Skorokhod topology), this convergence implies also convergence of the corresponding point process.

To keep the discussion simple we ignored so far (in this section) the existence of the internal degree of freedom $y_j$.   To incorporate that, one may consider a variant of the above argument, with a `decorated' version of the function $K(x;\omega)$, for which values of the variable $y$ are associated to the discontinuities of the function $K$. (Although this is not true for the full range of c\`adl\`ag functions, for $K\in \H$ the collection of discontinuities varies  continuously with  $K$ in the Skorokhod topology which is employed here.)   The above argument applies then mutatis mutandis.

This may be a place to note that the strategy employed here for the proof of convergence of the point process has its roots in the proof by A. Lenard of convergence of the Gibbs states of one-dimensional Coulomb systems, based on the analysis of the corresponding electric field ensemble~\cite{L}.     For the proof of symmetry breaking, we employ  additional  arguments which were introduced in~\cite{AM,AGL}.

\vspace{4mm}

\noindent{\bf Mutual singularity of limiting measures at different values of $\theta$; reconstruction of the phase $\Phi(\omega) $}

The above construction of the limiting measures for the point process in
 $\mathcal{T}$, proceeds through the construction of a limiting measure for the variable $K\in \H$.    In terms of this variable, the quantity we are after is
\be
\Phi(\omega) = e^{i2\pi K(0;\omega)} \, .
\ee
This expression however does not suffice:  for our purpose it is essential to establish that the phase $\Phi $  can be evaluated as a measurable (and thus quasi local) function of the \emph{point configuration}.    That may not be obvious at first sight since a shift of the function by a constant:  $K(x) \mapsto K(x) + C$ changes the value of $\Phi$,  without affecting the point configuration (that is the set of discontinuities of $K$).
However, the information provided by Theorem~\ref{thm:average} is of help here.

Using the coboundary relation \eqref{coboundary}, for any $R>0$
\begin{eqnarray}   \label{KfromQ}
 K(0;\omega) & = & - \frac{1}{R} \int_{0}^R [K(u;\omega)-  K(0;\omega)] \, du
 \  +  \     \frac{1}{R} \int_{0}^R K(u;\omega) \, du   \nonumber \\[2ex]
 & = & -\frac{1}{R} \int_{0}^R q^{-1} Q(u;\omega) \, du\  +  \     \frac{1}{R} \int_{0}^R  K(u;\omega) \, du
\nonumber  \\[2ex]
 & = &     \sum_{j:\, 0<x_j<R}  \left|1- \frac{x_j}{R}\right| \ -\ \frac{R}{2\lambda}
 \  +  \     \frac{1}{R} \int_{0}^R  K(u;\omega) \, du
 \, .
\end{eqnarray}
Summing the probability bound \eqref{eq:average} we find that for any of the finite-volume Gibbs measures:
\be  \label{vanishing}
 \mu_{[n_1,n_2]}^{(\beta,W,\theta)}
		\left(\left\{ \sup_{\ell  \ge R} \left| \frac{1}{\ell} \int_{0}^\ell K(u;\omega) \, du \,  \ - \
1/\ell  \right| \ge \delta  \right\}\right) \ \le \ \frac{\widetilde C }
		{ 1-e^{-\alpha \  \delta^2 }} \  \   e^{-\alpha  \  \delta^2  \  R }  \, ,
\ee
(where the insignificant  $1/\ell $ correction allows  to relate the maximum of $|K(x;\omega)|$ within intervals of length $\lambda$ to the end-point values).
 It now easily follows (by considering the implications of \eqref{eq:average}  and then applying the Borel Cantelli lemma) that the following limit converges almost surely with respect to any probability measure which is an accumulation point of the finite-volume $\theta$ Gibbs states (the statement denoted here by ${`\theta - a.s.'}$ )
\be  \label{eq:reconstruction}
\Phi(\omega)  \  \stackrel{\theta- a.s.}{:=} \   \lim_{R \to \infty} \   \exp \left\{ i 2\pi \left[ \sum_{j:\, 0<x_j<R}  \left|1- \frac{x_j}{R}\right| \ -\ \frac{R}{2\lambda} \right] \right\}
\ee
The construction also guarantees that $\Phi(\omega)$ satisfies
\be
\Phi(\omega) \  \stackrel{\theta - a.s.}{=} \   e^{i2\pi \theta}
\ee
and by implication also the relations which were  claimed in  \eqref{eq:phimaps}.
\end{proof}

\noindent {Remarks:}
\begin{enumerate}
\item
Extending the above considerations, one may obtain from equations \eqref{KfromQ} and \eqref{vanishing} an algorithm for the almost sure reconstruction, with respect to the limiting measure,  of  the function $K$ from the location of its discontinuities.
A fully deterministic quasi-local reconstruction is not possible, as the observation made above shows.
\item The argument used above bypasses the question of translation invariance.   Assuming it, or a slightly weaker regularity statement~\cite{mooreschmidt}, the reconstruction of  $K(x;\omega)$    could be done using Proposition~\ref{prop:cocycle} through the combination of tightness of the charge cocycle, which follows from the bounds of Theorem~\ref{thm:tightness},  and the normalization of the reference level which is implied by Theorem~\ref{thm:average}.
\end{enumerate}

\medskip

 \section{The one-dimensional case}\label{1Dsection}

It is now left to derive the bounds which enable the above analysis.
To introduce some of the ideas which are used in the proof in a somewhat simpler context,  let us first consider the analogous question for strictly one-dimensional  Coulomb systems.  Thus we consider particles of Coulomb charge $-1$ $(q=1)$ on a line in the presence of a uniform positive background charge, of  density given by the Lebesgue measure ($\rho=1$).

 We start with a bound on the probability of a uniformly large charge imbalance in such 1D systems with appended fixed charges at the endpoints.
 We denote here by $\{x\}=x-\lfloor x\rfloor$ the fractional part of $x\in \R$.

\begin{lemma}\label{lemma1}
For a  1D jellium system of $R =   \lfloor r \rfloor$ particles in an interval $[0,r]$,
with an `external' charge  $\gamma\in \N$ affixed
 at $x= 0$ and charge $-(\gamma+\{r\})$ affixed at $x=r$, the Gibbs probability $\mathbb{P}$ satisfies,
at any $\gamma \in (0,r)$:
	\begin{equation}\label{lemma1bound}
 	\mathbb{P}\left(\min_{0\le x\le r} \K(x) \ge \gamma\right) \le \frac{e^{-\frac{1}{2}\beta(\gamma^2-1)({R}-\gamma)}}
 		{e^{-(2{R}+\gamma\ln \gamma)}}\,.
 	\end{equation}
\end{lemma}
	
\begin{proof}   We recall that in one dimension the energy of a neutral configuration is given simply by
 \begin{equation} \label{1Denergy}
 		U(\omega) = \frac{1}{2} \int_0^r \K(x;\omega)^2 \dd x \, .
 \end{equation}
With the affixed boundary charges, we have that
 $\K(x;\omega)=\gamma+x -|\{i:\thinspace x_i<x\}|.$

 Let $\triangle$ be the $R$-simplex $\{0 <x_1<\cdots<x_{R} < r\}$
 	of all possible particle configurations. The edges of $\triangle$
 in the axes directions have length $r$.
 Consider the two subsets $\triangle^+$ consisting of all configurations for which $\K(x;\omega)\ge \gamma$ uniformly in $x$,
 and $\triangle^-$ consisting of all configurations for which:
 \begin{itemize}
 \item There are $\gamma+1$ particles in unit interval $[0,1]$, hence $\K(1) = \K(0)+ 1 - (\gamma +1) = 0$).
 \item There is exactly one particle per unit interval $\{(k-1,k):2\le k\le R -\gamma\}$, hence $\K(k) = 0$ for those $k$.
 \item There are no particles in $[R-\gamma,r]$, and the charge imbalance increases linearly from $0$ to $\gamma +\{r\}$ in this interval.
 \end{itemize}

For the Gibbs measure with the boundary charges as specified above, we have:
\begin{eqnarray}\label{1Dbound}
\nonumber\mathbb{P}\left(\min_{0\le x\le R} \K(x)=\gamma\right) &\le&
\frac{\int_{\triangle^+}   \exp[-\frac{1}{2}\beta\int \K(x;\omega)^2 dx]\ \mu(d\omega)}
{\int_{\triangle^-} \exp[-\frac{1}{2}\beta\int \K(x;\omega)^2 dx]\ \mu(d\omega)}\,.\\
&\leq& \exp \left(- \inf_{\omega\in\triangle^+}\beta U(\omega) + \sup_{\omega\in \triangle^-}\beta U(\omega) \right) \frac{\mu(\triangle^+)}{\mu(\triangle^-)}\, ,
	\end{eqnarray}
with $\mu(d\omega)$ the Lebesgue measure on $\triangle$.     We shall now estimate the energy and volume factors of the RHS
of \eqref{1Dbound} separately.

\paragraph{Energy estimate:} \mbox{ }

For $\omega \in \triangle^+$ we have (see Figure 3)
\be
\K(x;\omega) \geq \begin{cases}
    \gamma+\{x\},& \quad x \in [0, 1]\cup[R, r),\\
    \gamma,&\quad x\in [1, R],
	 \end{cases}
\ee
thus
\be\label{highenergybound}
 2 U( \omega)\ge A+\gamma^2(R-1)+ B\, ,
\ee
with $A=\int_{0}^{1} (\gamma+\{x\})^2 dx$ and $B=\int_{R}^r (\gamma+\{x\})^2 dx$.

For $\omega \in \triangle^-$,
\be  |\K(x)| \leq \begin{cases}
	 \gamma +\{x\}, & \quad x \in [0, 1]\cup [R, r),\\
	 1, &\quad x \in [1, R-\gamma+1], \\
               \gamma, &\quad x \in [R-\gamma+1, R] \, ,
					\end{cases}
\ee
whence
\be
2 U( \omega)\leq
A+(1)^2(R-\gamma)+\gamma^2(\gamma-1)+B \, .
\ee
Putting this together, we get the following bound on the ``improvement in energy'', i.e. its lowering, of configurations
in $\triangle^-$ in comparison to those in $\triangle^+$:
	\begin{equation} \label{eq:minimal-gap}
		\inf_{\triangle^+} U - \sup_{\triangle^-} U  \geq (R-\gamma) (\gamma ^2 - 1)\,.
	\end{equation}	

\begin{tikzpicture}[scale=.7]
 \path (0,0) coordinate (origin);
 \path (0,3) coordinate (o2);
 \path (6.25,3.25) coordinate (end2);
 \path (0,-1) coordinate (low);
 \path (6.25,4.5) coordinate (high);
 \path (10,0) coordinate (origin2);
 \path (10,3) coordinate (o22);
 \path (10,-1) coordinate (low2);
 \path (16.25,4.5) coordinate (high2);

 \draw (0,-2.2) node [right,text width=60mm]
    {\footnotesize Figure 3: The function $K(x)$ for configurations in $\triangle^+$ and $\triangle^-$.};

\draw [step=0.5,thin,gray!40] (low) grid (high);
 \draw[color=black,fill={yellow!50}] (origin)--(o2)--++(.5,.5)--++(0,-3.5) -- cycle;
 \draw (1,2) node [right] {$A$ energy};
 \draw [thick,->] (.8,2) --++ (-.5,0);
 \draw[fill={yellow!50}] (end2)--++(0,-3.25)--++(-.25,0)--++(0,3) -- cycle;
 \draw (5.4,1.6) node [left] {$B$ energy};
 \draw [thick,->] (5.6,1.6) --++ (.5,0);

 \draw [color=black,->] (origin) node [below, xshift=-.3cm] {$0$} --++(6.25,0) node [below, xshift=.2cm] {$r$};
 \draw (0,3) node [left] {$\gamma$};
 \draw [color=black,->] (low) -- ++(0,5.5) node [left] {$\K(x)$};
 \draw (3,-1) node [below] {$\omega\in \triangle^+$};

 \draw [blue] (origin)--(o2)--++(.8,.8)--++(0,-.5)--++(.3,.3)--++(0,-.5)--++(.5,.5)--++(0,-.5)
 --++(.7,.7)--++(0,-.5)--++(.6,.6)--++(0,-.5)--++(.5,.5)--++(0,-.5)
 --++(.4,.4)--++(0,-.5)--++(.8,.8)--++(0,-.5)--++(.3,.3)--++(0,-.5)
 --++(.6,.6)--++(0,-.5)--++(.5,.5)--++(0,-.5)--++(.1,.1)--++(0,-.5)--(end2)--++(0,-3.25);

 \draw [step=0.5,thin,gray!40] (low2) grid (high2);
 \draw[fill={yellow!50}] (origin2)--(o22)--++(.5,.5)--++(0,-3.5) -- cycle;
 \draw[fill={yellow!50}] (16.25, 3.25)--++(0,-3.25)--++(-.25,0)--++(0,3) -- cycle;

 \draw [->] (origin2) node [below, xshift=-.3cm] {$0$} --++(6.25,0) node [below, xshift=.2cm] {$r$};
 \draw (10,3) node [left] {$\gamma$};
 \draw [->] (low2) -- ++(0,5.5) node [left] {$\K(x)$};
 \draw (13,-1) node [below] {$\omega\in \triangle^-$};

 \fill[color=gray] (10.5,3) rectangle (13.5,0.5);
 \draw (12,2.8) node [below] {improved};
 \draw (12,1.8) node [below] {energy};

 \draw [blue] (origin2)--(o22)--++(.2,.2)--++(0,-.5)--++(.03,.03)--++(0,-.5)--++(.03,.03)--++(0,-.5)
 --++(.03,.03)--++(0,-.5)--++(.03,.03)--++(0,-.5)--++(.03,.03)--++(0,-.5)
 --++(.03,.03)--++(0,-.5)--++(.38,.38)--++(0,-.5)--++(.4,.4)--++(0,-.5)
 --++(.45,.45)--++(0,-.5)--++(.55,.55)--++(0,-.5)--++(.7,.7)--++(0,-.5)--(16.25,3.25)--(16.25,0);
 \end{tikzpicture}

\paragraph{Volume factors:}
	Applying the two-sided Stirling approximation
	\begin{equation*}
		\sqrt{2\pi n}(n/e)^{n} \leq n!\leq \sqrt{2\pi n}(n/e)^n e^{1/12}
	\end{equation*}
	we find that
	\begin{equation}\label{vol1}
		\mu(\triangle^+) \leq \frac{r^{R}}{{R}!}\le\frac{({R}+1)^{R}}{\sqrt{2\pi {R}}({R}/e)^{R}}
	\le \frac{e^{2{R}}}{\sqrt{2\pi {R}}}\,.
	\end{equation}
	and
	\begin{equation}\label{vol2}
		\mu(\triangle^-)  = \frac{1}{(\gamma +1)!} \geq \frac{e^{\gamma+1-1/12}}{\sqrt{2\pi(\gamma+1)}(\gamma+1)^{\gamma+1}}\,.
	\end{equation}

Combining Eqs.~\eqref{eq:minimal-gap}, \eqref{vol1}, and \eqref{vol2} we obtain
\begin{eqnarray}\label{volumecomparison}
\nonumber			\mathbb{P}\left(\min_{0\le x\le R} \K(x)=\gamma\right)
&\le& \frac{\mu(\triangle^+)}{\mu(\triangle^-)}
\exp[-\frac{\beta}{2}\left((\gamma^2-1)({R}-\gamma)\right)]\\
\nonumber&\le&\frac{e^{2{R}+(\gamma+1)\ln({\gamma+1})}}{e^{\gamma+1-1/12}}\times\exp[-\frac{\beta}{2}\left((\gamma^2-1)({R}-\gamma)\right)]\\
&\le&e^{2{R}+\gamma\ln\gamma}\times\exp[-\frac{\beta}{2}\left((\gamma^2-1)({R}-\gamma)\right)]
\end{eqnarray}
\end{proof}

We now consider the 1D jellium on $[L_1,L_2]$, with $[L_1<0<L_2]$. First, let us define two `crossing events' (analogous to $\triangle^\pm$ of the previous lemma) which play a key role in the sequel:

\begin{definition}
For $\gamma,\ell \in \N$ and $r>0$, let $\GG^+(\ell,r)$ be the set of configurations satisfying
\begin{enumerate}
		\item[(+a)] $\K(0)>3\gamma$ and $\K(x)>\gamma$ for $- \ell < x< r$
		\item[(b)] $\K(-\ell) = \gamma$
		\item[(c)] $\K(r-) > \gamma \geq \K(r+)$. Equivalently,
					$ \K(r-) = \gamma + \{r\}$ and $\K(r+) = \gamma + \{r\} - 1$.	
	\end{enumerate}
\end{definition}

\begin{definition}
	Let $\GG= \GG(\ell,r)$ be the set defined by the two conditions (b) and (c) above, but with (+a) replaced with
	\begin{itemize}
		\item[(a)] $\K(x)$ does not equal or cross the value $\gamma$  for $x\in[-\ell+1,R)$.
	\end{itemize}
\end{definition}
	Condition (a) of $\GG$ ensures that the sets $\GG(\ell,r)$ ($\ell \in \N$, $r>0$) are disjoint. In particular, suppose
$\mathcal{H}$ is the event that $K(x)\ge \gamma$ for at least one $x$-value both to the left and right of $x=0$.  Then $[-\ell,-\ell+1)$ and $[R,R+1)$ are
the first unit intervals to the right and left of the origin for which $K(x)$ ``crosses" $\gamma$. The events $\GG(\ell,r)$ is the decomposition of $\mathcal{H}$ into such events, thus
\be\label{H-decomp}
\mathcal{H}=\bigcup_{\ell,r} \GG(\ell,r)
\ee
Obviously, $\GG^+\subset \GG$. We will write $\GG^{(+)}$ when a statement applies to both $\GG$ and $\GG^+$.

\begin{theorem}\label{1Dthm}
	The probability distribution of the charge imbalance, $\K(x)$, in an overall neutral 1D Jellium in $[L_1,L_2]$, with $L_1<0<L_2 (\in \Z)$,
	satisfies:
	\begin{equation}
		\mathbb{P}\left(|\K(0)|>3\gamma\right) \ \le \  c_2 e^{-c_1\gamma^3} \, ,
	\end{equation}
	for some $c_1(\beta),c_2(\beta)>0$.
\end{theorem}

\begin{proof}
The strategy of proof is as follows: by symmetry, it is enough to consider the case $\K(0) \ge 3\gamma$ for some fixed $\gamma$.
Because of the boundary conditions $\K(L_1) = \K(L_2) =0$ (overall neutrality with no boundary charges), the charge imbalance
must cross the line $\K(x) = \gamma$ to the left and right of the origin. We group together configurations that
have the same crossing points. We may suppose without loss of generality that there are no particles at integer coordinates
and no two particles have the same coordinate, $x_k \notin \Z$ and $x_i <x_j$ for $i<j$.

Denote the simplex of configurations of $N=L_2-L_1$ particles by
\be\label{simplexdef}
 \triangle(N,[L_1,L_2]) := \{ (x_1,...,x_N) \in [L_1,L_2]^N:\ x_1<x_2< \cdots < x_{N} \}
\ee
and by $\int \dd \hat \omega$ the Lebesgue integration with respect to all $x$-variables except the $k$th one pinned at  $x_k = r$,
where $k=|L_1|+R+1-\gamma$ (recall that $R=\lfloor r\rfloor$).
Let also:
\begin{equation}\label{eq:densityE}
			p(\GG(\ell,r)) := \frac{\int_{\GG} \exp(- \beta U(\omega)) \dd \hat \omega}
			    {\int_{\triangle} \exp(- \beta U(\omega)) \dd \omega}\,.
	\end{equation}
Note that by the disjointness of the events $\GG(\ell,r)$ and \eqref{H-decomp},
$$\sum_{\ell= 1}^{|L_1|} \int_0^{L_2} \dd r\thinspace p(\GG(\ell,r)) =  \P{\mathcal{H}}$$
thus $p(\GG(\ell,r))/\P{\mathcal{H}}$ is a probability density.

It follows that
\begin{equation} \label{eq:conditioned}
	  \P{\K(0)>3\gamma}  \le \sum_{\ell= 2\gamma+1}^{|L_1|} \int_0^{L_2} \dd r\thinspace p(\GG^+ | \GG )p(\GG)  \leq
			\sup_{\ell,r} p(\GG^+ | \GG)\,.
\end{equation}
since $\{\K(0)>3\gamma\} = \cup_{\ell> 2\gamma,r>0}\GG^+(\ell,r)$. Here $p(\GG^+|\GG)$ is using the Radon-Nikodym derivative, $p(\GG^+|\GG) := p(\GG^+)/p(\GG)$
where $p(\GG^{+})$ is defined in the fashion of \eqref{eq:densityE}.

	The density $p(\GG^+\mid \GG)$ is equal to the probability of (+a) occurring
	on $[-\ell,r]$ with boundary charges as prescribed in (b) and (c); this is the so-called Markov property of 1D Coulomb systems. Thus $p(\GG^+\mid \GG)$ is in the form required by Lemma~\ref{lemma1}.

Note that $K(0)\ge 3\gamma$ implies
$\ell\ge 2\gamma+1$.
Suppose now that
$\gamma\ge\sqrt{1+\frac{6}{\beta}}$ which in turn implies $\frac{1}{2}\beta(\gamma^2-1)\ge 3$. A little algebra gives us
\begin{eqnarray}
\P{\K(0)>3\gamma}
\nonumber&\le&\sup_{R,\ell}\frac{e^{-\frac{1}{2}\beta(\gamma^2-1)({R+\ell}-\gamma)}}
 		{e^{-(2{(R+\ell)}+\gamma\ln\gamma)}}\\
\nonumber&\le& \sup_{R,\ell} e^{-c_1\gamma^2({R+\ell}-\gamma)}\\
&\le& e^{-c_1\gamma^3}
\end{eqnarray}
for some $c_1(\beta)>0$. We can then choose $c_2(\beta)>0$ large enough so that for all $\gamma$, $\P{\K(0)>3\gamma} \le c_2 e^{-c_1\gamma^3}$.
\end{proof}

Before moving on to the quasi-1D systems, let us give a refinement of the bound \eqref{highenergybound} which will be useful for proving an
analogous result to Theorem \ref{1Dthm} for the strip.

\begin{lemma}[Improved energy bound] \label{refinement_lemma}
	Let $\omega \in \GG^+$. If $n_k(\omega)$ is the number of particles in $[k,k+1]$ for \mbox{$k = -\ell,..., R-1$} and $n_R(\omega)$ is the
	number of particles in $[R,r)$, then $n_{-\ell}(\omega) = 0 $ and
	\begin{equation}\label{refinedbound}
		 \int_{-\ell}^r \K(x;\omega)^2 \dd x  \geq \gamma^2(\ell +r)+ \sum_{k= -\ell}^R \bigl(\gamma n_k(\omega)^2 + n_k(\omega)^3 / 3\bigr)\,.
	\end{equation}
\end{lemma}

\begin{proof}
	$n_{-\ell} = 0$ follows from $\K(-\ell +1) = \gamma + 1 - n_{-\ell}   >\gamma$ and the integrality of the involved quantities.
	More generally, $n_k \leq k+\ell$ and $\sum_{k=-l}^m n_k \le m+\ell$ for all $m\le R$.
	The energy estimate follows from the observation that any non-zero $n_k$ adds at least a triangle (see Figure 4) with
	side length $n_k$ to the field $\K(x),x\leq k$, on top of $\gamma$:
	\begin{eqnarray*}
		\K(k) = \gamma +(k+\ell) - \sum_{j<k}n_j \geq& \gamma + n_k.
	\end{eqnarray*}
	This gives additional factors in the energy
\be  \int_0^{n_k} 2 \gamma x \dd x + \int_0^{n_k} x^2 \dd x= \gamma n_k(\omega)^2 + n_k(\omega)^3 / 3. \qedhere \ee
\end{proof}

\begin{tikzpicture}[scale=.7]
 \draw (0,-2) node [right,text width=80mm]
    {\footnotesize Figure 4: Excesses in the energy of $\omega \in \GG^+$, as estimated in ~\eqref{refinedbound}.};
 Draw axes and labels
 \path (0,0) coordinate (origin);
 \path (0,1.5) coordinate (o2);
 \path (7.75,1.75) coordinate (end);
 \path (0,-1) coordinate (low);
 \path (7.75,5.5) coordinate (high);

 \draw [->] (origin) node [below, xshift=-.2cm] {$0$} --++(7.75,0) node [below, xshift=.2cm] {$R$};
 \draw (0,1.5) node [left] {$\gamma$};
 \draw [->] (low) -- ++(0,6.5) node [left] {$\K(x)$};
 \draw [step=0.5,thin,gray!40] (low) grid (high);

 \fill[color=gray] (.5,1.5) rectangle (6.5,0.5);
 \draw (3,1.5) node [below] {improved energy};

 \draw [fill={yellow!50}] (7,5) -- (7,2) -- (4,2) -- cycle;
 \draw [fill={yellow!50}] (4.5,5) -- (4.5,2.5) -- (2,2.5) -- cycle;
 \draw [fill={yellow!50}] (1.5,3) -- (1.5,2) -- (.5,2) -- cycle;
 \draw [fill={yellow!50}] (7.5,2.5) -- (7.5,1.5) -- (6.5,1.5) -- cycle;

 \draw (5.3,6.9) node [below] {additional};
 \draw (5.3,6.4) node [below] {improved energy};
 \draw [thick,->] (5.9,5.6) -- (6.4,3.6);
 \draw [thick,->] (4.4,5.6) -- (3.9,3.6);

 Draw \K(x)
 \draw [blue] (origin)--(o2)--++(1.8,1.8)--++(0,-.5)--++(.1,.1)--++(0,-.5)--++(2.76,2.76)--++(0,-.5)
 --++(.1,.1)--++(0,-.5)--++(.05,.05)--++(0,-.5)--++(.1,.1)--++(0,-.5)
 --++(.04,.04)--++(0,-.5)--++(2.09,2.09)--++(0,-.5)--++(.05,.05)--++(0,-.5)
 --++(.05,.05)--++(0,-.5)--++(.05,.05)--++(0,-.5)--++(.04,.04)--++(0,-.5)
 --++(.03,.03)--++(0,-.5)--++(.3,.3)--++(0,-.5)--++(.03,.03)--++(0,-.5)--(end)--++(0,-1.75);
 Draw \K(x) (2)
 \draw [blue] (origin)--(o2)--++(.02,.02)--++(0,-.5)--++(.03,.03)--++(0,-.5)--++(.04,.04)--++(0,-.5)
 --++(.04,.04)--++(0,-.5)--++(.55,.55)--++(0,-.5)--++(.56,.56)--++(0,-.5)
 --++(.57,.57)--++(0,-.5)--++(.50,.50)--++(0,-.5)--++(.4,.4)--++(0,-.5)
 --++(.45,.45)--++(0,-.5)--++(.55,.55)--++(0,-.5)--++(.7,.7)--++(0,-.5)
 --++(.55,.55)--++(0,-.5)--++(.45,.45)--++(0,-.5)--++(.45,.45)--++(0,-.5)
 --(end)--++(0,-1.75);
\end{tikzpicture}

\section{Tightness bounds}  \label{sec:tightness}
\label{replacementsection}

We now turn to the proof of Theorem~\ref{thm:tightness} for jellium in the tube $\mathcal{T}$.
 The strategy  is as  demonstrated in the 1D case.  However, the energy estimates are complicated
by the $V_2$-interaction. We denote by $V_2(\omega_1,\omega_2)$ the sum of the
$V_2$-interaction terms of pairs of distinct particles, one in $\omega_1$ and the other in $\omega_2$,  and by $V_2(\omega):=V_2(\omega,\omega)$ such a sum for a single configuration, omitting self interactions.
Following the decomposition~\eqref{kernel_expansion}, the energy of a configuration $\omega =(z_1,...,z_N)$
 is a sum of a 1D part and a short-range interaction energy.
\begin{equation} \label{eq:energy-decomposition}
 	U(\omega) = \frac{q^2}{2W}\int_{L_1}^{L_2} |\K(x;\omega)|^2 \dd x + q^2\sum_{1\leq j<k\leq N}
 			V_2(y_j,y_k;|x_j-x_k|) = U^1(\omega) + V_2(\omega). 	
 \end{equation}
Given $\ell>0$, $r>0$, the tube $\mathcal{T}$ splits into subsystems
\begin{equation} \label{eq:subsystems}
 	\Y=\Y_{\ell,r}:=[-\ell,r] \times \mathbb{D},\quad \L\r:=\mathcal{T} \backslash \Y.
\end{equation}
and the total energy naturally decomposes into
\begin{equation}\label{eq:semi-markov}
 	U(\omega)= U_{\L\r}(\omega_{\L\r}) + U_\Y(\omega_\Y) +  V_2(\omega_\Y,\omega_{\L\r}) \, .
\end{equation}

Note that in one dimension $V_2\equiv 0$ so that the last term on the RHS of
\eqref{eq:semi-markov} vanishes.  This yields the Markov property for the function $K(x;\omega)$ in one-dimensional Coulomb systems (where it plays the role of the electric field).

In the current setting, we must control
\begin{equation} \label{eq:v2terms}
	V_2(\omega_\Y) + V_2(\omega_\Y,\omega_{\L\r}) \, .
\end{equation}

Following are two useful perspectives on the arguments which are presented below.

\paragraph{An alternative viewpoint: Replacing particles with rods}
To analyze $V_2$-interactions we can employ the procedure of \emph{replacing} a particle at $z_k$ with a  \emph{rod}, or in dimension $d\geq 3$ a $\D$-shaped charge slab, of vanishing thickness
${\bf \bar x}_k= \{ z:\thinspace x=x_k\}$, in which the charge $(-q)$ is uniformly distributed with charge density
$-q W^{-1}\dd y$.    By \eqref{average} the `rod-particle' or `rod-rod` interaction is completely free of the term $V_2$, and thus when the particles in $\Y$ are replaced by rods, the rod's  distribution conditioned on the rest of the system is identical to that of a strictly one-dimensional system.
In this
setting,  one  will broaden the notion of a configuration of particles to a \emph{generalized} configuration
of particles and rods. The $x$-values will still be ordered, and the replacement $z_k\mapsto \mathbf{\bar{x}}_k$
corresponds to the map
\be
	\omega=(z_1,\ldots,z_{N}) \mapsto \omega^{\{{\bf \bar{x}_k}\}}=(z_1,\ldots,z_{k-1},
	{\bf \bar{x}}_k,z_k,\ldots,z_{N})\, .
\ee
Each replacement changes the energy by an amount equal to~\eqref{eq:v2terms}:
Thus, denoting by $
	\omega \mapsto \omega^\Y
$
(generalized) configuration in  which  all particles in $\Y$ were replaced by rods, we have \begin{equation} \label{eq:energychange}
	 U(\omega) - U( \omega^\Y) =  V_2(\omega_\Y) + V_2(\omega_\Y,\omega_{\L\r}).
\end{equation}

Therefore there are two equivalent points of view for the estimates of the next section: we can either think of
$V_2$-energy estimates for a standard system of point particles, or we can think of the estimate as bounding the  \emph{replacement
effect} on the energy incurred when all particles in $\Y$ are replaced by rods.

\paragraph{Jellium as a system of unbounded spins}  One may also regard the system discussed here  as one of unbounded spins, with additional internal degrees of freedom.  There is however,  a significant  difference, which does not make the  existing bounds for such systems with `superstable interactions' ~\cite{ruelle,ruelle-spins,lebowitz-presutti} directly applicable.   The ``spins'' are the charge imbalances at integer $x$  coordinates, $\K(k;\omega)$, $k\in \Z$.
The internal degrees of freedom are the exact locations, including $y$ coordinates, of particles in $[k,k+1) \times \D$.

The $V_2$-interaction between two disjoint regions $X$ and $Y$  is
bounded below by $- q^2 g(\dist(X,Y)) n(X) n(Y)$, with $n(X)$ the number of particles in $X$. This is reminiscent of \emph{lower regularity}
as defined in \cite{ruelle}. \emph{Upper regularity}, i.e., a (pointwise) converse inequality, does not hold since $V_2 = \infty$ is possible.
However, a suitable substitute can be obtained through the zero mean of $V_2$ over $y$ (applying the Jensen inequality).

The representation~\eqref{eq:energy-decomposition} together with $V_2(\omega) \geq - q^2 \sum_{j<k} n_k(\omega) n_j(\omega) g(k-j-1)$ suggests that
an inequality of the form
\begin{equation}
	U(\omega) \geq \sum_k [ A \K(k;\omega)^2 - B]
\end{equation}
for suitable $A,B>0$ might hold, i.e., the system of spins might be \emph{superstable}.

This is really close to situations considered in~\cite{ruelle,ruelle-spins}.
In fact, the proof of Theorem~\ref{thm:tightness} is close in spirit to the proofs in those papers. One complication arising
for jellium is that we cannot simply decrease one spin $\K(k;\omega)$ without affecting other spins:
point charges cannot be removed without affecting an infinite change in the energy, but they can be  moved from one place to another.  In other words,  the overall neutrality fixes the total number of particles,
in contrast to the grand canonical setting of~\cite{ruelle}.   The procedure employed here circumvents this difficulty: as in the 1D case, we look not at an individual large spin but at a selected subsystem $[-\ell,r]\times \D= \Y$, inside of which charges are rearranged leaving the charge distribution outside unchanged.
A technical difficulty here, shared with the systems of~\cite{ruelle, ruelle-spins}, is that making
some spin smaller may actually increase its interaction energy with other spins.

In the rest of the section, we shall write $\mathbb{P}$ instead of $\mu_{[n_1,n_2]}^{(\beta,W,\theta)}$, and without loss of generality, we will
assume $\theta =0$.

\subsection{$V_2$-energy estimates} \label{sec:v2estimates}

We would like bounds on what a replacement does to the total energy and to the
Boltzmann weight of a configuration $\omega$. The first step towards this end is the zero average
property~\eqref{average}. Together with Jensen's inequality applied to the exponential function,
it yields
\begin{equation} \label{eq:rodjensen}
	e^{-\beta U(\omega^{\{{\bf \bar{x}_k}\}})}
		\leq  \frac{1}{W} \int_{\D} \ e^{-\beta U(\omega)} dy_k.
\end{equation}
Our goal for the rest of this subsection will be to prove a sort of converse to~\eqref{eq:rodjensen}.

Let $\ell\in \N$, $r>0$ and $\Y\subset \mathcal{T}$ as in~\eqref{eq:subsystems}.
Let $R= \lfloor r/\lambda \rfloor$. We divide $\Y$
into \emph{cells}
\be
\Y_k = [k\lambda,(k+1)\lambda)\times \D \ \text{ for }\ k=-\ell,\ldots, R-1\quad \text{and}\quad\Y_R = [R\lambda,r]\times \D,
\ee
and denote by $n_k(\omega)$  the number of particles in $\Y_k$.

We start by noting that $V_2$ is ``lower regular'' in the sense that it can be bounded below in terms of the ($\omega$-dependent) particle numbers $n_k(\omega)$.  In this statement a role is played by:

\begin{lemma}[Lower bound on $ V_2(\omega_\Y)$] \label{lemma:y-estimate}
For all $\omega$, the $V_2$-energy satisfies
	\begin{equation}\label{eq:y-estimate}
		 V_2(\omega_\Y) \geq - q^2C \sum_{k=-\ell}^R n_k(\omega_\Y)^2
	\end{equation}
	for some constant $C = C(W) >0$.
\end{lemma}

\begin{proof}
By the assumption which was made in \eqref{lowerbound},
the interaction $V_2$ between particles, in cells $\Y_{k_1}$, $\Y_{k_2}$  (not necessarily distinct) which are distance $D\lambda\geq 0$ from each other, is  bounded below by
\be  \label{nnbound}
	 - q^2 g(D \lambda) \, n_{k_1}(\omega) \, n_{k_2}(\omega) \, ,
\ee
with $g(u)$ the non-negative monotone  decreasing function which appears there.
The integrability assumption on $g$ implies that the sum over integer multiples of $\lambda$ is finite,
	$\sum_{D\in \N} g(D\lambda) <\infty$.
	
Let $k_0$ be the value of $k$ such that $n_k(\omega) \leq n_{k_0}(\omega)$ for all $k=-\ell,\ldots, R$.
	The interaction between particles in $\Y_k$ (or $\Y_R=[R \lambda,r]\times \D$ if $k=R$)
	with themselves and with others is lower bounded by
	\begin{multline}
		- q^2 g(0)n_{k_0}(\omega)^2/2  - q^2 \sum_{k=-\ell}^{k_0-1} g([k_0-k-1]\lambda)n_{k_0}(\omega) n_k(\omega)
			- q^2 \sum_{k=k_0+1}^R g([k-k_0-1]\lambda) n_{k_0}(\omega) n_k(\omega) \\
			\geq - q^2 \left(\frac{5}{2} \, g(0) + 2 \sum_{D = 1}^\infty g(D\lambda)\right) n_{k_0}(\omega)^2 =:
			 -q^2C n_{k_0}(\omega)^2.
	\end{multline}
	Next, we can find $k_1$ maximizing $n_k(\omega)$ among $k\neq k_0$, and estimate
	interactions between the remaining cells, $k= -\ell, \ldots, R$, $k\neq k_0$, in a strictly
	analogous way.  Iterating the procedure until all cells have been taken care of, yields the lemma.
\end{proof}

Next, we need to control the interaction between  $\Y$ and the rest of the tube ($\L\r$).
 Obviously,  as in \eqref{nnbound}, the interaction between particles in, e.g., $\Y_{-\ell}$ and the left subsystem may be bounded below by
\be
	-  q^2 \sum_{D=0}^{L_1-\ell-1} g(D\lambda) \, n_{-\ell}(\omega) \, n_{-\ell-D-1}(\omega) \, .
\ee
However, for long tubes, with $L_1 >>1 $, this bound can in principle become very negative.  That happens in case there is an accumulation of  particles of  $\L\r$ close to $\Y$.  We will need to control the frequency of occurrence of such ``irregular'' configurations.

Let $\gamma \in \N$. Let $\GG\subset\Omega$ be defined as in Section \ref{1Dsection} with suitable $y$ degrees of freedom
added. Recall that for $\omega \in \GG$,
\begin{equation}
  \K(-\ell;\omega) = \gamma, \qquad \K(r-;\omega) > \gamma \geq \K(r+;\omega).
\end{equation}
We will call a configuration $\omega \in \GG$ \emph{regular} if its charge imbalance
 is well-behaved outside $\Y$ in the
sense that
\be\label{niceomega}
	|\K(x;\omega)| \le \gamma+\text{dist}(x,[-\ell,R]) \text{ for }x\notin [-\ell,r].
\ee
The set of regular configurations will be denoted here $\GG_\mathrm{reg} \subset \GG$.

\noindent \emph{Remark:} On $\GG$, we automatically have that $\K(x;\omega)\ge -\gamma-|x+\ell|$ for $x<-\ell$ and that $\K(x;\omega)\le \gamma+|x-\{r\}|$ for $x>R$ (see the regions
corresponding to the dotted triangular regions in Figure 5), thus the condition \eqref{niceomega}
really amounts to restricting $\K(x;\omega)$ from entering the shaded triangular regions of Figure~5 below.

\begin{tikzpicture}[scale=.5]
\draw (15,0) node [right,text width=55mm]
    {\footnotesize Figure 5:  Regular and irregular configurations $\omega\in \GG$, in the sense of \eqref{niceomega}.
 For regular configurations the graph of $K$ does not enter the shaded region.};
\draw (15,-3) node [right,text width=55mm]
    {\footnotesize Within $[-\ell,R+1]$, the function $K$ can cross $\gamma$ only in the first and last unit intervals, marked by black strips.};

 \path (0,0) coordinate (origin);
 \path (0,5) coordinate (lhigh);
 \path (14,0) coordinate (end);
 \path (0,-5) coordinate (llow);
 \path (14,5) coordinate (rhigh);
 \path (14,-5) coordinate (rlow);

 \draw [step=0.5,thin,gray!40] (llow) grid (rhigh);
 \draw [->] (origin) node [left] {$0$} --(end) node [below, xshift=.2cm] {$x$};
 \draw [->] (llow) -- (lhigh) node [left] {$\K(x)$};
 \draw [thick, black] (3.5,0) node [below, xshift=-1mm] {$-l$} -- (10.5,0) node [below] {${R}$};
 \draw [dashed,black!40] (0,1.5) node [left, black] {$\gamma$} -- (14,1.5);
 \draw [dashed,black!40] (0,-1.5) node [left, black] {$-\gamma$} -- (14,-1.5);

 \draw [fill, gray!40] (lhigh)--++(3.5,-3.5)--++(0,3.5)--cycle;
 \draw [fill, gray!40] (rlow)--++(-3.5,3.5)--++(0,-3.5)--cycle;
 \draw [dotted, thick, gray] (rhigh)--++(-3.25,-3.25)--++(0,3.25)--cycle;
 \draw [dotted, thick, gray] (llow)--++(3.5,3.5)--++(0,-3.5)--cycle;

 \draw [blue] (0,3.5)--++(.03,.03)--++(0,-.5)--++(.03,.03)--++(0,-.5)--++(.03,.03)--++(0,-.5)
 --++(.03,.03)--++(0,-.5)--++(.02,.02)--++(0,-.5)--++(.02,.02)--++(0,-.5)
 --++(.03,.03)--++(0,-.5)--++(.02,.02)--++(0,-.5)--++(.02,.02)--++(0,-.5)
 --++(.03,.03)--++(0,-.5)--++(.03,.03)--++(0,-.5)--++(3.21,3.21);

 \draw [blue] (10.75,1.75)--++(.45,.45)--++(0,-.5)--++(.15,.15)--++(0,-.5)
 --++(.1,.1)--++(0,-.5)--++(.25,.25)--++(0,-.5)--++(.3,.3)--++(0,-.5)--++(.25,.25)--++(0,-.5)--++(.3,.3)--++(0,-.5)
 --++(.1,.1)--++(0,-.5)--++(.25,.25)--++(0,-.5)--++(.3,.3)--++(0,-.5)--++(.25,.25)--++(0,-.5)--++(.35,.35)--++(0,-.5)
 --++(.2,.2);

  \draw [red] (0,2.5)--++(.73,.73)--++(0,-.5)--++(.39,.39)--++(0,-.5)--++(.54,.54)--++(0,-.5)
 --++(.28,.28)--++(0,-.5)--++(.5,.5)--++(0,-.5)--++(.46,.46)--++(0,-.5)
 --++(.27,.27)--++(0,-.5)--++(.23,.23)--++(0,-.5)--++(.1,.1)--++(0,-.5);

  \draw [red] (10.75,1.75)--++(0,-.5)
 --++(.08,.08)--++(0,-.5)--++(.07,.07)  --++  (.13,.13)  --++(0,-.5)--++(.05,.05)--++(0,-.5)
 --++(.05,.05)--++(0,-.5)--++(.03,.03)--++(0,-.5)--++(.05,.05)--++(0,-.5)--++(.02,.02)--++(0,-.5)--++(.08,.08)--++(0,-.5)
 --++(.03,.03)--++(0,-.5)--++(.04,.04)--++(0,-.5)--++(.75,.75)--++(0,-.5)--++(.29,.29)--++(0,-.5)--++(.25,.25)--++(0,-.5)
 --++(.16,.16)--++(0,-.5)--++(.05,.05)--++(0,-.5)--++(.12,.12)--++(0,-.5)--++(.1,.1)--++(0,-.5)--++(.9,.9);

  \draw [purple] (3.5,1.5)--++(.7,.7)--++(0,-.5)--++(.45,.45)--++(0,-.5)--++(.55,.55)--++(0,-.5)
 --++(.4,.4)--++(0,-.5)--++(.5,.5)--++(0,-.5)--++(.7,.7)--++(0,-.5)--++(.45,.45)--++(0,-.5)--++(.55,.55)--++(0,-.5)
 --++(.4,.4)--++(0,-.5)--++(.5,.5)--++(0,-.5)--++(.7,.7)--++(0,-.5)--++(.45,.45)--++(0,-.5)
 --++(.55,.55)--++(0,-.5)--++(.35,.35)--++(0,-.5);

  \draw [purple] (3.5,1.5)--++(.3,.3)--++(0,-.5)--++(.15,.15)--++(0,-.5)--++(.2,.2)--++(0,-.5)
 --++(.4,.4)--++(0,-.5)--++(.2,.2)--++(0,-.5)--++(.7,.7)--++(0,-.5)--++(.45,.45)--++(0,-.5)--++(.55,.55)--++(0,-.5)
 --++(.4,.4)--++(0,-.5)--++(.5,.5)--++(0,-.5)--++(.65,.65)--++(0,-.5)--++(.45,.45)--++(0,-.5)--++(.65,.65)
 --++(0,-.5)--++(.45,.45)--++(0,-.5)--++(1.2,1.2);

 \draw [very thick](3.5,1.5) --++ (.5,0);
 \draw [very thick](10.5,1.5) --++ (.5,0);
 \end{tikzpicture}
\vspace{1mm}

For configurations in $\GG_\mathrm{reg}$ we can
lower bound the $V_2$-interaction and the total energy change due to a replacement:

\begin{lemma}[Lower bound for $V_2(\omega_\Y,\omega_{\L\r})$] \label{lemma:y-reg}
	For all $\omega \in \GG_\mathrm{reg}$, the $V_2$-energy satisfies
	\begin{equation}
		V_2(\omega_\Y,\omega_{\L\r}) \geq - q^2C \gamma \max_{ - \ell \leq k \leq R} n_k(\omega)
	\end{equation}
	for some constant $C= C(W)>0$.
\end{lemma}

\begin{proof}
	For $k=-\ell,...,R$ let $n_k(\omega)$ be the particle numbers introduced above. We extend the definition
	to $k\in \Z$, $L_1 \leq k\lambda \leq L_2-\lambda$ in the obvious way. Call $n_r(\omega)$ the number of particles in $(r,(R+1)\lambda)\times \D$.
	Using the monotonicity of  $g(u)$, the interaction between a particle $z \in \Y$ with
	the particles from $\omega$ in $x<-\ell$ is lower bounded by 	
	\begin{equation}
		 - q^2\sum_{k=0}^{L_1 - \ell -1} g([k+x+\ell]\lambda) \, n_{-\ell - k-1}(\omega) \  \geq \
			 - 2 q^2 \sum_{k=0}^\infty g([k+x+\ell]\lambda)  \, .
	\end{equation}
	The inequality is obtained as follows: sum by parts in order to rewrite the sum with
	differences of $g$ and sums of particle numbers, use that $g$ is decreasing and $\omega$ regular,
	and sum by parts again. Thus
	the total interaction of all particles in $\Y$ with the left substrip is bounded by
	\begin{align*}
		- 2q^2 \,  \sum_{p=-\ell}^R n_p(\omega)  \sum_{k=0}^\infty g([p+\ell+k]\lambda)
		\  \geq  \    - q^2 \left(\max_{k=-\ell,...,R}
				n_k(\omega)\right) \sum_{k=0}^\infty 2(k+1) g(k\lambda).
	\end{align*}
	The interaction with the right tube $x>r$ can be bounded in a similar way. An additional summand
	$\gamma$ will appear because for regular configurations, there may be as many as $2\gamma$ particles
	accumulated near $x = r$ (see Fig.~4 above).
\end{proof}

\begin{lemma}[Coupling lemma]\label{lemma:coupling}
	For all $\omega_\Y$ (a configuration of
	particles inside $\Y$),
	\begin{equation}\label{eq:coupling}
		\int_{\GG} e^{-\beta U(\omega)} \dd \omega_{\L\r}
			\leq e^{C (\gamma^{3/2} + n_\mathrm{max}(\omega)^{3/2} + \gamma
			 n_\mathrm{max}(\omega))}\int_{\GG_\mathrm{reg}}e^{-\beta U(\omega)} 			
			\dd \omega_{\L\r}
	\end{equation}
for some constant $C = C(\beta, W)>0$ and $n_\mathrm{\max}(\omega) = \max_{- \ell \leq k \leq R} n_k(\omega)$. The integrals
	are over configurations $\omega_{\L\r}$ where $$\omega:=\omega_{\L\r} \cup \omega_\Y \in \GG_{(\mathrm{reg})}.$$
\end{lemma}

In somewhat loose notation, Eq.~\eqref{eq:coupling} may be thought of
as a lower bound
for the probability of regular configurations, given that they are in $\GG$ and that
the configuration inside $\Y$ is $\omega_\Y$
\begin{equation}
	\P{\GG_\mathrm{reg}\mid \GG, \omega_\Y}  \geq \exp\Bigl(- C (\gamma^{3/2} + n_\mathrm{max}(\omega_\Y)^{3/2} + \gamma
			 n_\mathrm{max} (\omega_\Y))\Bigr)
\end{equation}

In this sense Lemma~\ref{lemma:coupling} says that regular configurations have high enough probability.
We defer the proof of this lemma to Section \ref{sec:coupling}.

Using Lemmas \ref{lemma:y-estimate},\ref{lemma:y-reg}, and \ref{lemma:coupling}, we can now formulate the ``converse'' to Eq.~\eqref{eq:rodjensen}. Note that
$U(\omega^\Y)$ is the energy of the system with all particles in $\Y$ replaced with
rods.
\begin{lemma}[Replacement lemma] \label{lemma:repl}
	For all $\omega_\Y$,
	\begin{equation} \label{eq:repl}
		\int_\GG e^{-\beta U(\omega)} \dd \omega_{\L\r}
		 \leq   e^{C (\gamma^{3/2} + \gamma
			 n_\mathrm{max}(\omega_\Y)+\sum_{k=-\ell}^R n_k(\omega_\Y)^2)} \int_\GG e^{-\beta U(\omega^\Y)} \dd \omega_{\L\r},
	\end{equation}
	for some constant $C=C(\beta,W)$.
\end{lemma}
Thus, on average, the Boltzmann weight before replacement is smaller than the Boltzmann weight after
replacement, up to some (controllable) function of $\omega_\Y$.

\subsection{Proof of Theorem~\ref{thm:tightness}}

The strategy of proof is exactly the same as in the 1D case.
Let $\gamma, -L_1,L_2 \in \lambda \N$. We want to estimate the probability of the event $\K(0;\omega)\geq 3\gamma$.
Let $\GG^{(+)}(\ell,r)$ be events defined as in Section~\ref{1Dsection}. Note that since the events
were defined in terms of charge imbalances which depend on $x$-coordinates solely, we can extend
the definitions to the quasi-1D case by adding the $y$ degrees of freedom.
The definition of conditional densities $p(\GG^{(+)})$, $p(\GG^{(+)}|\GG)$ is extended in the natural way as well.
By the same argument as in Section~\ref{1Dsection},
\begin{equation}\label{eq:mainbound}
	\P{ \K(0;\omega) \geq 3\gamma } \leq \sup_{\ell,r} p(\GG^+(\ell,r)|\GG(\ell,r))
\end{equation}

On $\GG_{\text{reg}}$,
Lemmas \ref{lemma:y-estimate} and \ref{lemma:y-reg} give us control over the $V_2$-interactions, enough so that the ``1D-portion" of the energy implies
the desired exponential decay of the RHS of \eqref{eq:mainbound}. However, to deduce the exponential decay for all of $\GG$, we need Lemma \ref{lemma:repl} (which combines Lemmas \ref{lemma:y-estimate},\ref{lemma:y-reg}, and \ref{lemma:coupling}).

Recall that $U(\omega^\Y) = U_\Y^1(\omega_\Y) + U_{\L\r}(\omega_{\L\r})$
is the energy of the system with particles in $\Y$ replaced with rods.
By Lemma~\ref{lemma:repl},
\begin{equation}
	\int_{\GG^+(\ell,r)} e^{-\beta U(\omega)} \,\dd \hat \omega \leq
	\int_{\GG^+(\ell,r)} e^{-\beta U(\omega^\Y)}
			 e^{C(\gamma^{3/2} + \gamma
			 n_\mathrm{max}(\omega_\Y)+\sum_{k=-\ell}^R n_k(\omega_\Y)^2)} \,\dd \hat \omega
\end{equation}
where we recall that $\dd \hat \omega$ means Lebesgue measure of all variables except the $x$-variable pinned at $x=r$,
and by Jensen and Eq.~\eqref{average}
\begin{equation*}
	\int_{\GG(\ell,r)} \exp(-\beta U(\omega)) \,\dd \hat \omega
	 \geq \int_{\GG(\ell,r)} \exp(- \beta U(\omega^{\Y})) \,\dd \hat \omega.
\end{equation*}
It follows that
\begin{equation*}
	p(\GG^+|\GG) \leq \frac{\int_{\GG^+} \exp(- \beta U(\omega^\Y))
			\exp[C(\gamma^{3/2}+\gamma n_{\max}(\omega)+\sum_{k=-\ell}^R n_k(\omega_\Y)^2] \dd \hat \omega
        }		
		{\int_{\GG} \exp(- \beta U(\omega^\Y)) \dd \hat \omega}.	
\end{equation*}
Note that this upper bound is independent of $L_1$ and $L_2$.

Since in ${U(\omega^\Y)}$ there are no interactions between $\Y$ and $\L\r$, we have again a Markov property,
and the estimates reduce to estimates for the finite tube $\Y$. We can proceed as in Lemma~\ref{lemma1},
using Lemma~\ref{refinement_lemma} in order to take care of the extra factor
$\exp[C(\gamma^{3/2}+\gamma n_{\max}(\omega_\Y)+\sum_{k=-\ell}^R n_k(\omega_\Y)^2]$.

\subsection{Regular configurations have positive probability} \label{sec:coupling}

In this section we prove Lemma~\ref{lemma:coupling}. The strategy of proof is to define a map
\begin{equation}
	\GG \to \GG_\mathrm{reg}, \quad \omega \mapsto \omega'
\end{equation}
by shifting some of the particles in $\L$ and $\r$ (the left and right substrips in the complement of $\Y$) away from $\Y$, and to substitute
$\int_\GG \dd \omega$ by $\int_{\GG_\mathrm{reg}} \dd\omega'$. Complications arise because
the map is not one-to-one and has a non-trivial Jacobian, resulting in a compression of phase space, i.e.,
entropy loss.  However regular configurations have lower (1D) energy, and this energetic improvement is enough
to compensate for the entropy loss. \\

\noindent \emph{Remark.} The astute reader will notice the parallel between the above map and the
comparison (in the 1D case) of $\triangle^+$ and $\triangle^-$. Configurations in $\triangle^-$
are energetically more favorable but have smaller Lebesgue volume $\mu(\triangle^-)$.\\

Fix $\gamma,\ell\in\N$ and $r>0$. Let $R= \lfloor r /\lambda \rfloor$.
For $\omega \in \GG$, denote $\omega_\Y$ , $\omega_\L$ and $\omega_\r$ the projections
onto $\Y$, $\L$ and $\r$.  To simplify matters, let us focus on $\r$ and
pretend first that $\ell = L_1$ (which really cannot happen since $\K(-\ell;\omega) = \gamma > 0 = \K(L_1;\omega)$).
 There are $N_\r = L_2 - R - 1 + \gamma $ particles in $\r$.
Write $\omega_\r = \{z_1,...,z_{L_2-R-1+\gamma}\}$ with particles labeled from left to right,
$x_j <x_{j+1}$. We observe that if for all $k$
\begin{equation} \label{eq:regular-particle}
	x_k-r \geq (k-\gamma-1)\lambda/2,
\end{equation}
then $\omega \in \GG_\mathrm{reg}$. We will call particles \emph{regular} if they satisfy
 Eq.~\eqref{eq:regular-particle}, \emph{irregular} otherwise. Thus regularity of all particles implies regularity
of the configuration. The first $\gamma +1$ particles are  always regular.

 For $\omega \in \GG$, let $\omega'\in \GG$ be such that  $\omega'_\Y = \omega_\Y$  (no changes in $\Y$)
and $\omega'_\r$ the collection of points $\{z'_k\}$ with
\begin{equation}
	y'_k = y_k,\qquad x'_k = \begin{cases}
					x_k,& \quad \text{regular particle}, \\
					(R + k - \gamma)\lambda + \frac{x_k - r}{(k-\gamma - 1)/2}, &\quad
						\text{irregular}.
	                  \end{cases}
\end{equation}
 The idea behind the map is the following: for fixed charge imbalance $\K(r) \approx \gamma$,
 the 1D part of the energy in $\r$ is minimal if $\gamma$ particles accumulate near $x = r$
 and the remaining $L_2 - R - 1$ particles occupy each one of the ``equilibrium'' cells
 $\Y_{R+1}, ..., \Y_{L_2-1}$. The map $\omega \mapsto \omega'$ simply shifts
 an irregular particle closer to its equilibrium position, thereby decreasing the (1D) energy.

\begin{lemma}[Energy estimates]\label{lemma:energy-estimates}
	Let $I_\mathrm{irr}(\omega) \subset \{\gamma +2,...,N_\r\}$ be the set of irregular particle labels.
	The map $\GG \ni \omega \mapsto \omega'\in \GG_\mathrm{reg}$ decreases the 1D-energy by an amount which is at least
	\begin{equation} \label{eq:1Denergy}
		U^1(\omega) - U^1(\omega') \geq \frac{q^2 \lambda}{W}
			\sum_{k\in I_\mathrm{irr}(\omega)}\frac{ (k-\gamma)^2 - 1}{8}.
	\end{equation}
	Furthermore,
	\begin{align}
		 V_2(\omega_\Y,\omega_\r) & \geq - q^2C n_\mathrm{max}(\omega) \bigl(|I_\mathrm{irr}(\omega)|
				+ \gamma +1) \label{eq:vyr} \\
		 V_2(\omega_\r) & \geq - q^2C \sum_{k\in I_\mathrm{irr}(\omega)}(k+1) \label{eq:vrr}
	\end{align}
for some constant $C = C(W)>0$.	
\end{lemma}

Note that we do not evaluate $V_2(\omega) - V_2(\omega')$. Instead we estimate directly $V_2(\omega)$. Jensen's
inequality will take care of $V_2(\omega')$.

\begin{proof}[Proof of Lemma~\ref{lemma:energy-estimates}]
	\emph{The 1D-energy term:}   We shift irregular particles successively by the order of their subscript, starting with the highest one.
    In this
	way one obtains a sequence $\omega^{(n)}$ starting at $\omega^{(0)} = \omega$.
	The shift $x_k\to x'_k$ increases the particle imbalance in $[x_k,x'_k)$ by $1$
	and leaves it unchanged
	 elsewhere. This decreases the energy by an amount
	\begin{align}
		\notag U^1(\omega^{(n)}) - U^1(\omega^{(n-1)}) & = \frac{q^2}{2 W} \int_{x_k}^{x'_k}
			\Bigl[ \K(x;\omega^{(n)})^2 -  \Bigl( \K(x;\omega^{(n)}) +1 \Bigr)^2 \Bigr] \dd x \\
		\notag	&= - \frac{q^2}{W}\int_{x_k}^{x'_k} \Bigl[ \K(x_k;\omega) + \frac{1}{2}\Bigr] \dd x \\
			& \geq \frac{q^2 \lambda}{8 W}[ (k-\gamma)^2 - 1]
	\end{align}
	where we have used that the integrand is bounded above by an affine function with slope $1/\lambda$,
	end value $\leq 1/2$,
	integrated over an interval of length $x'_k - x_k \geq (k-\gamma)\lambda/2$.
	The sum of these bounds yields~\eqref{eq:1Denergy}.

\begin{tikzpicture}[scale=.5]
\draw (12,-2) node [right,text width=65mm]
    {\footnotesize Figure 6:  A possible energy improvement for an irregular
	 $\omega\in \GG$, which is obtained through the displacement of a particle from $x=a$  to $x=b$.   The modified $\K(x)$ is described by the dotted line.};

 \path (0,0) coordinate (origin);
 \path (0,3.25) coordinate (lhigh);
 \path (10.5,0) coordinate (end);
 \path (0,-3.5) coordinate (llow);
 \path (10.5,3.25) coordinate (rhigh);
 \path (10.5,-3.5) coordinate (rlow);

 \draw [step=0.5,thin,gray!40] (llow) grid (rhigh);
 \draw (origin) -- (end);
 \draw [->] (llow) -- (lhigh);
 \draw [thick, black] (0,0) node [below, xshift=1mm] {$-l$} -- (7.25,0) node [below] {$r$};
 \draw [dashed,black!40] (0,1.5) node [left, black] {$\gamma$} -- (10.5,1.5);
 \draw [dashed,black!40] (0,-1.5) node [left, black] {$-\gamma$} -- (10.5,-1.5);

 \draw [fill, gray!40] (9,-3.5)--++(-2,2)--++(0,-2)--cycle;;

  \draw [red] (7.25,1.75)--++(0,-.5)
 --++(.08,.08)--++(0,-.5)--++(.07,.07)  --++  (.13,.13)  --++(0,-.5)--++(.05,.05)--++(0,-.5)
 --++(.05,.05)--++(0,-.5)--++(.03,.03)--++(0,-.5)--++(.05,.05)--++(0,-.5)--++(.02,.02)--++(0,-.5)--++(.08,.08)--++(0,-.5)
 --++(.03,.03)--++(0,-.5)--++(.04,.04)--++(0,-.5);

 \draw [red, fill, yellow!50] (7.88,-3.12)--++(2.62,2.62)--++(0,.5)--++(-2.62,-2.62)--cycle;
 \draw [red] (7.88,-2.62)--++(0,-.5)--++(2.62,2.62);
 \draw [red, thick, dotted] (10.5,0)--++(-2.62,-2.62);

 \draw [red] (0,1.5)--++(.7,.7)--++(0,-.5)--++(.45,.45)--++(0,-.5)--++(.55,.55)--++(0,-.5)
 --++(.4,.4)--++(0,-.5)--++(.5,.5)--++(0,-.5)--++(.7,.7)--++(0,-.5)--++(.45,.45)--++(0,-.5)--++(.55,.55)--++(0,-.5)
 --++(.4,.4)--++(0,-.5)--++(.5,.5)--++(0,-.5)--++(.7,.7)--++(0,-.5)--++(.45,.45)--++(0,-.5)
 --++(.55,.55)--++(0,-.5)--++(.35,.35)--++(0,-.5);

 \draw [very thick](0,1.5) --++ (.5,0);
 \draw [very thick](7,1.5) --++ (.5,0);
 \draw (7.81,0) node [above] {\footnotesize $a$} circle (.06);
 \draw (end) node [above, xshift=-.5mm] {\footnotesize $b$} circle (.06);
\end{tikzpicture}
		
	\emph{$V_2$-energy inside $\r$:}  \  The individual terms $V_2$ are bounded below by  $- q^2 g(0)$.
	 Since there are
	$\sum_{k=1}^{|I_\mathrm{irr}|-1} k $ pairs of irregular particles, we get
	\be
		\sum_{j,k\in I_\mathrm{irr}\atop j < k} V_2(y_j,y_k;|x_j-x_k|)\geq  - q^2 g(0)
			\sum_{k\in I_\mathrm{irr}} k.
	\ee
	The interaction between a given irregular particle $z_k$ and the regular particles in $\r$
	is lower bounded by
	\begin{equation}
		 - q^2 \sum_{j\ \text{reg}} g(|x_j - x_k|\lambda)\geq - (k-1) q^2 g(0) - q^2 \sum_{j>k+1} g((j-k)\lambda/2).
	\end{equation}
	It follows that
	\begin{equation}
		\sum_{j\ \mathrm{reg}\atop k\ \mathrm{irr}} V_2(y_j,y_k;|x_j-x_k|)
		 \geq - q^2 \sum_{k\ \text{irr.}} \Bigl( g(0)k + \sum_{n=1}^\infty g(n\lambda/2)\Bigr)
	\end{equation}
	
	\emph{$V_2$-interaction between $\Y$ and $\r$:} \  Eq.~\eqref{eq:vyr} holds provided
	 $C$ is large enough so that
	\begin{equation}
		\sum_{m=0}^\infty \sum_{n=0}^\infty g(m\lambda+[n/2]\lambda) <C. \qedhere
	\end{equation}
\end{proof}

\begin{lemma}\label{lemma:7.6}
For all $\omega,\gamma$ and $\ell, r, L_1,L_2$,
	\begin{align}
		\notag &\int_{\D^{N_\r}} \exp(- \beta U(\omega)) \dd y_1 \cdots \dd y_{N_\r} \\
		\notag & \qquad \leq C'' \exp\bigl( C' ( \gamma^{3/2} + n_{\max}(\omega)^{3/2} + \gamma n_{\max}(\omega))
				 \bigr)
				\exp\left( - C \sum_{k \in I_\mathrm{irr}(\omega)} (k-\gamma)^2)\right) \\
			& \qquad \qquad \times 	
			\int_{\D^{N_\r}} \exp(- \beta U(\omega')) \dd y_1 \cdots \dd y_{N_\r},
	\end{align}
for some constants $C= C(\beta,W), C'= C'(\beta,W), C''= C''(\beta,W)$.
\end{lemma}

\begin{proof}
	\begin{align}
		\notag \int_{\D^{N_\r}} e^{-\beta U(\omega)} \dd {\boldsymbol y} &
			\leq e^{ - \beta [ U^1(\omega) - U^1(\omega') + \inf_{\boldsymbol{y}} (V_2(\omega_\r)
			+ V_2(\omega_\Y,\omega_\r))]}
				\int_{\D^{N_\r}} e^{-\beta U^1(\omega)} \dd \boldsymbol{y} \\
		& \leq e^{ - \beta [ U^1(\omega) - U^1(\omega') + \inf_{\boldsymbol{y}} (V_2(\omega_\r)
			+ V_2(\omega_\Y,\omega_\r))]} \int_{\D^{N_\r}} e^{-\beta U(\omega')} \dd \boldsymbol{y}
	\end{align}
	We might think of this inequality as a result of a three step procedure:
	(a) Replace particles with rods.
	The replacement error is quantified by Eqs.~\eqref{eq:vrr} and~\eqref{eq:vyr}.
	(b) Shift irregular rods away from $\Y$. The energy decrease is given by Eq.~\eqref{eq:1Denergy}.
	(c) Replace rods with particle averages using Jensen's inequality.
	
	Next, by Lemma~\ref{lemma:energy-estimates},
letting $\eta=k-\gamma$
	\begin{eqnarray}\label{eqn:3/2-power}
		\nonumber &&U^1(\omega) - U^1(\omega') + \inf_{\boldsymbol{y}} (V_2(\omega_\r)
			+ V_2(\omega_\Y,\omega_\r))\\
			 &\geq& \sum_{\eta:\thinspace (\eta + \gamma) \thinspace \in I_\mathrm{irr}}
			 \Bigl[ \frac{q^2 \lambda}{8 W}\eta^2
			- C (\gamma + \eta + n_\mathrm{max} +1) \Bigr] - C n_\mathrm{max}(\omega) (\gamma +1)
	\end{eqnarray}
	The $C\eta$ is dominated by $\eta^2$ except for small $\eta$. The $\gamma$ term inside
	the sum is dominated by $\eta^2$ except for $\eta^2 \lesssim \const \gamma$, whence a negative
	contribution of the order of $\gamma^{3/2}$, and similarly for $n_\mathrm{max}(\omega)$. Thus the RHS of \eqref{eqn:3/2-power} is greater
than $$C_1 \sum_{k \in I_\mathrm{irr}(\omega)} (k-\gamma)^2
				- C_2( \gamma^{3/2} + n_\mathrm{max}(\omega)^{3/2} + \gamma n_\mathrm{max}(\omega))$$
for suitable
	constants $C_1 = C_1(\beta,W)$, $C_2= C_2(\beta,W) $ (at this point the constants depend on $\beta$ only through their dependence on $q$; see the remarks concerning dimensionless parameters in Section \ref{sec:q1D jellium}).
\end{proof}

We are now equipped for the proof of Lemma~\ref{lemma:coupling}:

\begin{proof}[Proof of Lemma~\ref{lemma:coupling}]
	Let $\gamma, \ell, r$ be fixed. As in previous proofs, we will deal only with the region $\r$ ($\L$ is dealt with analogously).

Define
	the map $\omega\mapsto \omega'$ from $\GG$ to $\GG_\mathrm{reg}$ as above. Let $I \subset \{\gamma +2,...,
	N_\r\}$. Conditioned on the event that $I_\mathrm{irr}(\omega) = I$, the regularizing map
	is injective with the Jacobian
	\begin{equation}
		\left|\frac{\dd \omega'}{\dd \omega} \right| = \prod_{k\in I} \frac{2}{k-\gamma - 1}
			= \exp\left (- \sum_{k\in I} \ln [(k-\gamma - 1)/2]\right). 		
	\end{equation}
	Thus, for suitable constants,
	\begin{equation}
		\int_{I_\mathrm{irr}(\omega) = I} e^{-\beta U(\omega)} \dd \omega_{\r} \label{eq:Kint}
		 \leq C'' e^{-A(I)} \int_{\GG_\mathrm{reg}} e^{-\beta U(\omega')} \dd \omega'_{\r},
	\end{equation}
	with
	\begin{equation}
		A(I) := \beta C \sum_{k\in I} (k-\gamma)^2
			- \beta C'(\gamma^{3/2} + n_\mathrm{max}^{3/2} + \gamma n_{\max})
			- C'\sum_{k\in I} \ln [(k-\gamma - 1)/2] \, , 		
	\end{equation}
where the last sum   comes from the Jacobian.  We obtain a lower bound on $A(I)$ by dropping the last sum and adjusting the constants $C$ and $C'$. It follows that 	
\begin{equation}
		\sum_{I\subset \{\gamma +2,...,
	N_\r\}} \exp( - A(I))
			 \leq e^{C'(\gamma^{3/2} + n_\mathrm{max}^{3/2} + \gamma n_{\max})}
							\prod_{k=1}^\infty (1+e^{-C k^2}).
	\end{equation}
	Summing Eq.~\eqref{eq:Kint} over $I$ yields Lemma~\ref{lemma:coupling}, up to the simplification
	$\ell = L_1$.

	In the general case $\ell <L_1$, the regularizing map is extended by defining its action
	on particles in $\L$ in a similar way: particles too close to $\Y$ get shifted away. 	
	The energy estimates of Lemma~\ref{lemma:energy-estimates} extend in the natural way.
	From there one can proceed as in the fictitious case $\ell = L_1$.
\end{proof}

\section{Convergence of the volume-averages of $K(x;\omega)$}\label{sec:averages}

In this section, we prove Theorem \ref{thm:average}.
In order to acclimate ourselves, let us first prove the 1D case via a shift coupling for fixed $\delta>0$, $-L_1,L_2\in\N$, and ${r}$ as in Theorem \ref{thm:average}. As before, without loss of generality, we assume $\theta =0$.

Recall that $x_i$ denotes the position of the $i$th particle with $i<j$ implying that $x_i<x_j$ almost surely. The coupling $\omega\mapsto\tilde{\omega}$ is defined by moving the particles in the
following manner:
\begin{itemize}
\item If $x_i<2\delta$ then $\tilde x_i= x_i/2$
\item If $2\delta<x_i<{r}$ then $\tilde x_i = x_i-\delta$
\end{itemize}

By symmetry, we may think of Theorem \ref{thm:average} as a statement concerning the exponential decay (in ${r}$) of the probability of
$$
A_\delta := \{\omega:\int_0^{r} \K(x;\omega) \, dx \ge \delta {r} \}.
$$
To prove this exponential decay, we use the observation that using the above coupling,
for $2\delta<x<{r}$,
\be\label{eq:shiftcoupling}
\K(x;\omega) -\delta = \K(x-\delta;\tilde{\omega})\, .
\ee

\begin{proposition}\label{prop:1D shift coupling}
	Consider a one-dimensional Coulomb system. For all $\delta>0$ and $c<1$, there is a constant ${r}_0(\delta,c)$ such that ${r}\ge {r}_0$ implies
	\begin{align}\label{shift energy diff}
		\P{A_\delta}&\le e^{-c\beta\delta^2{r}}.
	\end{align}
\end{proposition}

\begin{proof}
	If we define the set
\be\label{def:F}
		\F=\{\omega:|\K(k;\omega)|\le {r}^{2/5}-1\text{ for each integer }k\in[0,{r}]\},
\ee
then by Theorem \ref{1Dthm}
we have $$\P{\F^c}\le ({r}+1)c_2(\beta) e^{-c_1(\beta) {r}^{6/5}}.$$
	Note that if $\omega\in \F$ then \mbox{$|\K(x;\omega)|\le {r}^{2/5}$} and $|\K(x;\tilde{\omega})|\le {r}^{2/5}+\delta$ for
	 all $x\in[0,{r}]$.
	Thus for $\omega\in A_\delta\cap \F$:
	\be\label{interior A_delta}
		\int_{2\delta}^{{r}} \K(x;\omega)\, dx \ge \delta {r} - 2\delta {r}^{2/5},
	\ee
	and using \eqref{eq:shiftcoupling},
	\begin{align}\label{1d energy change}
		\nonumber \int_0^{r} \K(x;\omega)^2 - \K(x;\tilde{\omega})^2 \, dx &\ge \int_{2\delta}^{{r}} (2\delta \K(x;\omega) -
			 \delta^2)\, dx- \int_{[0,\delta]\cup[{r}-\delta,{r}]}\K(x;\tilde{\omega})^2 \, dx.\\
		\nonumber &\ge 2\delta(\delta {r} - 2\delta {r}^{2/5}) - \delta^2({r}-2\delta)-
			 \int_{[0,\delta]\cup[{r}-\delta,{r}]}\K(x;\tilde{\omega})^2 \, dx\\
		\nonumber &\ge \delta^2 ({r}+2\delta - 4{r}^{2/5}) - 2\delta ({r}^{4/5}+2\delta {r}^{2/5} + \delta^2)\\
			&= \delta^2 ({r} - 8{r}^{2/5} - 2\delta^{-1}{r}^{4/5})
	\end{align}

	To get a bound on the probability we now only need a bound on the change in the Jacobian or ``volume-factor" caused by
	going from $\omega$ to $\tilde{\omega}$.  The only place where
	volume is lost is in the first step of the coupling when $[0,2\delta]$ is mapped to $[0,\delta]$.  For $\omega \in \F$
	there are at most $2{r}^{2/5}+2\delta$ particles in $[0,2\delta]$, thus
	the volume factor is bounded above by $e^{(2{r}^{2/5}+2\delta)\log 2}$. Altogether we have
	\be
		\P{A_\delta\cap \F^c}+\P{A_\delta\cap \F}\le ({r}+1)c_2 e^{-c_1 {r}^{6/5}} + e^{-\beta\delta^2[{r} - (8+2\log
		 2/(\beta\delta^2)){r}^{2/5} - 2\delta^{-1}{r}^{4/5} -(2\log 2)(\beta\delta)^{-1}]}
	\ee
	from which \eqref{shift energy diff} readily follows.
\end{proof}

For the general quasi-1D setting, the coupling $\omega\mapsto\tilde{\omega}$ maps $x_i\mapsto \tilde x_i$ just as above and leaves the $y_i$'s unchanged ($\tilde y_i=y_i$).
We need to adjust the proof of Proposition \ref{prop:1D shift coupling} in two places.  The first trivial adjustment is to bound the probability of $\F^c$ with Theorem \ref{thm:tightness}
(replacing the 1D bound of Theorem \ref{1Dthm}). The non-trivial adjustment is to show that on $\F$, any increase in the short-range $V_2$-interaction energy caused by the coupling
is insignificant compared to the decrease in 1D-energy given in \eqref{1d energy change}.

\begin{proof}[Proof of Theorem \ref{thm:average}]
Let $\F$ be as in \eqref{def:F} and define the `regular set' $\F_\mathrm{reg}$ similarly to the definition of $\GG_\mathrm{reg}$ in Section \ref{sec:v2estimates}, with ${r}^{2/5}$ playing the role of $\gamma$.
	By Lemma \ref{lemma:coupling} we have that
	\be\label{Freg}
		\P{\F_\mathrm{reg}}>e^{-c {r}^{4/5}}\P{\F}
	\ee
	for some $c>0.$ We now condition on $\F_\mathrm{reg}$ and bound
$\int\exp(- \beta V_2(\omega) )\,\dd \boldsymbol{y} \Big/\int\exp(- \beta V_2(\tilde \omega) )\,\dd \boldsymbol{y}$
on this event.

Recall (from Section \ref{replacementsection}) that $\Y_{a,b}=[a,b]\times\D$ and that $\omega^{\Y_{a,b}}$ is $\omega$ with all particles in $\Y_{a,b}$
	replaced with rods. If there are $k$ particles in the region
	$\Y_{0,\delta}$ (for $\tilde \omega$) and the region $\Y_{0,2\delta}$ (for $\omega$), and ${\boldsymbol y}$ denotes the
	 vector of $y$-coordinates of those particles, then for every $\omega\in\F$
	\begin{align}\label{eq:ubd}
		\notag &\int_{\D^k} \exp(- \beta V_2(\omega))\dd \boldsymbol{y} \\
		\notag &\qquad \leq \exp(- \beta \inf_{\boldsymbol{y}\in \D^k}[V_2(\omega) - V_2(\tilde \omega^{\Y_{0,\delta}})] )
				\int_{\D^k} \exp(-\beta V_2(\tilde \omega^{\Y_{0,\delta}})) \dd \boldsymbol{y} \\
			& \qquad \leq \exp\Bigl( - \beta \inf_{\boldsymbol{y}\in \D^k}[ V_2(\omega)- V_2(\omega^{\Y_{0,2\delta}})]
			- \beta [V_2(\omega^{\Y_{0,2\delta}}) - V_2(\tilde \omega^{\Y_{0,\delta}})]\Bigr)
			\int_{\D^k}  \exp(-\beta V_2(\tilde \omega)) \dd \boldsymbol{y},
	\end{align}
	where we have compared the integrals in the last two lines using Jensen's inequality applied to rod replacements. We have slightly abused notation
	in that the infimum over ${\boldsymbol y}\in \D^k$ is at different $x$-values for
	the coupled configurations $\tilde\omega$ and $\omega$.

 Note that the coupling does not affect $V_2$-interactions between two particles that are both in $\Y_{2\delta, {r}}$ (pre-coupling).
Therefore, on the event $\F_\mathrm{reg}$, the affected $V_2$-interaction can be lower bounded by
	\begin{equation}
		V_2(\omega) - V_2(\omega_{\L\r}) - V_2(\omega_{\Y\backslash \Y_{0,2\delta}})=
		V_2(\omega_\Y,\omega_{\L\r}) + V_2(\omega_{\Y_{0,\delta}}) + V_2(\omega_{\Y_{0,\delta}},\omega_{\Y\backslash
			\Y_{0,\delta}} ) \geq - c {r}^{4/5}.
	\end{equation}

If we can now show that conditioned on $\F_\mathrm{reg}$,
\be \label{bound1}
V_2(\omega^{\Y_{0,2\delta}})-V_2(\omega) \le C {r}^{4/5}
\ee and
\be \label{bound2}
V_2(\tilde{\omega}^{\Y_{0,\delta}})-V_2(\omega^{\Y_{0,2\delta}}) \le C {r}^{4/5},
\ee then the 1D decrease in energy caused by the coupling (see \eqref{1d energy change}) overwhelms the coefficients of the right-hand sides of both \eqref{eq:ubd} and \eqref{Freg} which would prove the theorem. Let us now show \eqref{bound1} and \eqref{bound2}.

To bound
	$V_2(\omega^{\Y_{0,2\delta}})-V_2(\omega)$ on $\F_\mathrm{reg}$, note that there are at most $2{r}^{2/5}+2\delta$ particles in $\Y_{0,2\delta}$. Replacing the role of $\max n_k$ with $2{r}^{2/5}+\delta$ and the role of $\gamma$ with ${r}^{2/5}$ in Lemma \ref{lemma:y-reg}, we obtain
\be
V_2(\omega^{\Y_{0,2\delta}})-V_2(\omega) < C ({r}^{4/5}+\delta {r}^{2/5})
\ee
 for some $C>0$.
	
To bound $V_2(\tilde{\omega}^{\Y_{0,\delta}})-V_2(\omega^{\Y_{0,2\delta}})$, first recall that $\omega_{\Y_{0,{r}}}$ and $\omega_{\Y^c_{0,{r}}}$ denote the sets of particles of $\omega$ that are in $\Y_{0,{r}}$ and its complement, respectively.
	Using \eqref{eq:energychange}, since the $V_2$-interactions within $\Y_{0,{r}}$ are the same for both $\tilde{\omega}^{\Y_{0,\delta}}$ and $\omega^{\Y_{0,2\delta}}$, we have
	\begin{align}\nonumber V_2(\tilde{\omega}^{\Y_{0,\delta}})-V_2(\omega^{\Y_{0,2\delta}})&=V_2(\tilde{\omega}^{\Y_{0,\delta}}_{\Y_{0,{r}}},\tilde{\omega}^{\Y_{0,\delta}}_{\Y_{0,{r}}^c} )-V_2(\omega^{\Y_{0,2\delta}}_{\Y_{0,{r}}},\omega^{\Y_{0,2\delta}}_{\Y_{0,{r}}^c}).
	\end{align}

By \eqref{upperbound} we have that for each $\delta>0$ there exists $C(\delta)>0$ such that
\be	
V_2(z_1,z_2)\le C g(|x_1-x_2|)
\ee
 whenever $|x_1-x_2|\ge \delta$. The condition $|x_1-x_2|\ge \delta$ is satisfied by
	$z_1\in\tilde{\omega}^{\Y_{0,\delta}}_{\Y_{0,{r}}}$ and $z_2\in \tilde{\omega}^{\Y_{0,\delta}}_{\Y_{0,{r}}^c}$, so again following the proof of
	Lemma \ref{lemma:y-reg}, we get that
	$V_2(\omega^{\Y_{0,2\delta}})-V_2(\omega) < C {r}^{4/5}$ for some $C>0$.
\end{proof}

\section{Discussion }\label{sec:discussion}

The results presented here confirm a conjecture which was stated in \cite{AGL}.   Let us point out some related questions which were not addressed in this work.

\begin{enumerate}
\item
It may be worth stressing that the length $\lambda  \equiv (\rho W)^{-1}$, with which translation symmetry breaking is proven here for shifts by $u\notin \lambda \Z$, does not correspond to the interparticle spacing.   The difference between the two length scales shows up only for values greater than $1$ of the dimensionless parameter $W/\lambda^{(d-1)}$, but it becomes very  pronounced when $W\gg \lambda^{(d-1)}$.
In particular, this means that the symmetry breaking proven here does not correspond to the phenomenon of Wigner lattice.

\item
A question which was not addressed is whether the symmetry breaking stops at the level which is proven here, e.g. whether for each $\theta$ the Gibbs measures $\mu^{(\beta, W, \theta)}_{n_1,n_2}$ have a unique limit, for $n_1,n_2 \to \infty$.    The answer would be negative if, for instance,  for at least certain values of $W$ the system is in a lattice like state wrapped on the cylinder.
\item
Let us express here the conjecture that the answer to the above question is positive.  More explicitly:   we expect that for  each $\theta$ there is a unique limiting state.   Furthermore we expect this state to be  not only invariant under the shift $T_{\lambda}$ (which would be implied by the uniqueness) but also ergodic with respect to it.   In particular this would mean that the states do not admit any further cyclic decomposition; i.e., the symmetry breaking stops at the level which is proven here.    Can that be shown in a brief argument?

\item  Adding a comment to  the above question:  in case of the cylinder we do not expect the  symmetries of rotation in the compactified dimensions to be  broken.

 \item
 As is the case for point processes which are not too singular,
 the $\theta$-states discussed can be characterized through their correlation functions, $\{\rho_n(z_1,\ldots,z_n)\}$. Thus, the state's non-invariance under shifts implies that at least some of these correlation functions (if not all)  are not shift invariant.   In \cite{JLS} it was shown that in narrow enough strips translation symmetry breaking occurs already at the level of the one-point density function.   Does such a statement  extend to the entire regime covered by the non-perturbative argument presented here?
\item
Other examples of particles with  Coulomb potential include two component systems, of particles of charges $\pm q$ (and not necessarily equal masses). For such systems on a line, there is no translation symmetry breaking, but there is a related phenomenon of phase non-uniqueness~\cite{AM}.   In essence, in one dimension the fractional part of a charge placed at the boundary cannot be screened by any readjustment of the integer charges.  Does  this phenomenon also persist to the quasi one-dimensional Coulomb systems?   We conjecture that the answer is affirmative even though   the rigidity of the $1D$ Coulomb interaction is somewhat `softened' in the quasi one-dimensional extension.
\end{enumerate}

\section*{Acknowledgements}
S. Jansen wishes to thank E. H. Lieb and Princeton University for making possible the stay during which this work was initiated.   Some of the work was done when M. Aizenman was visiting IHES (Bures-sur-Yvette) and the Center for Complex Systems at the Weizmann Institute of Science.   He wishes to thank both institutions for their hospitality. Finally, P. Jung thanks UCLA and IPAM for their hospitality during which some of this work was done.

\end{document}